\documentclass[aps,prb,twocolumn,showpacs,superscriptaddress,10pt]{revtex4-1}\begin{tiny}\end{tiny}
\usepackage{longtable}
\usepackage{amsfonts}
\usepackage[dvipdfmx]{graphicx}
\usepackage{rotating}
\usepackage{hyperref}
\usepackage{dcolumn}
\newcolumntype{d}{D{.}{.}{3.5}}

\begin{document}

\title{Spin susceptibility and electron-phonon coupling
of two-dimensional materials
by range-separated hybrid density functionals:
Case study of Li$_x$ZrNCl}

\author{Bet\"{u}l Pamuk}
% \email{betul.pamuk@impmc.upmc.fr}
 \affiliation{CNRS, UMR 7590 and Sorbonne Universit\'{e}s, UPMC Univ Paris 06, IMPMC - Institut de Min\'{e}ralogie, de Physique des Mat\'{e}riaux, et de Cosmochimie, 4 place Jussieu, F-75005, Paris, France}

\author{Jacopo Baima}
 \affiliation{Dipartimento di Chimica and Centre of Excellence NIS (Nanostructured Interfaces and Surfaces), Universit\`{a} di Torino, via P. Giuria 5, I-10125 Turin, Italy}

\author{Roberto Dovesi}
 \affiliation{Dipartimento di Chimica and Centre of Excellence NIS (Nanostructured Interfaces and Surfaces), Universit\`{a} di Torino, via P. Giuria 5, I-10125 Turin, Italy}

\author{Matteo Calandra}
 \email{matteo.calandra@impmc.upmc.fr}
 \affiliation{CNRS, UMR 7590 and Sorbonne Universit\'{e}s, UPMC Univ Paris 06, IMPMC - Institut de Min\'{e}ralogie, de Physique des Mat\'{e}riaux, et de Cosmochimie, 4 place Jussieu, F-75005, Paris, France}

\author{Francesco Mauri}
 \email{francesco.mauri@impmc.upmc.fr}
 \affiliation{Dipartimento di Fisica, Universit\`{a} di Roma La Sapienza,
Piazzale Aldo Moro 5, I-00185 Roma, Italy}

\date{\today}

%\begin{spacing}{2}

\begin{abstract}
We investigate the capability of density functional theory (DFT) to appropriately describe the spin susceptibility, $\chi_s$, and
the intervalley electron-phonon coupling in Li$_x$ZrNCl.
At low doping, Li$_x$ZrNCl behaves as a two-dimensional two-valley electron gas, with parabolic bands.
In such a system, $\chi_s$ increases with decreasing doping because of the electron-electron interaction.
We show that DFT with local functionals (LDA/GGA) is not capable of reproducing this behavior.
The use of exact exchange in Hartree-Fock (HF) or in DFT hybrid functionals enhances $\chi_s$. HF, B3LYP, and PBE0 approaches overestimate $\chi_s$, whereas
the range-separated HSE06 functional leads to  results similar to those obtained in
the random phase approximation (RPA) applied to a two-valley two-spin electron gas.
Within HF, Li$_x$ZrNCl
is  even unstable towards a ferromagnetic state for $x<0.16$.
The intervalley phonons induce an imbalance in the valley occupation that can be viewed as the effect of a 
pseudomagnetic field.
Thus, similarly to what happens for $\chi_s$, the electron-phonon coupling of intervalley phonons is enhanced by the electron-electron interaction.
Only  hybrid DFT functionals capture such an enhancement
and the HSE06 functional reproduces the RPA results presented in
M. Calandra {\it et al.} [Phys. Rev. Lett. {\bf 114}, 077001 (2015)].
These results imply that the description of the susceptibility
and electron-phonon coupling with a range-separated hybrid functional
would be important also in other two-dimensional weakly doped semiconductors,
such as transition-metal dichalcogenides and graphene.

\end{abstract}

%\pacs{29.85.-c}
\maketitle

%       \newpage
%       \tableofcontents
%       \newpage        

\section{Introduction}

Low doping of layered multivalley semiconductors is a field of intense research in nanotechnology
and superconductivity \cite{Saito2015, Ye2015}.
Fairly high T$_c$ values have been reported with doped two-dimensional semiconductors,
such as transition metal dichalcogenides \cite{Novoselov2005,Xu2014,Zhang2014,Ye2012,Ye2015},
ternary transition-metal dinitrides \cite{Gregory1998}, and
cloronitrides \cite{Yamanaka1996,Yamanaka1998};
the doping of which can be achieved and controlled by
intercalation \cite{Yamanaka1996,Yamanaka1998,Taguchi2006,Takano2008,Takano2008b,Yamanaka2009}
or field effect \cite{Novoselov2005,Xu2014,Ye2010,Kasahara2011,Thomas2014,Saito2015}.

Weakly doped two-dimensional, and quasi-two-dimensional (2D)
semiconductors composed
of weakly interacting layers stacked along the $z$-direction,
behave very differently than their 3D counterparts.
In 3D semiconductors, with parabolic bands, the density of states, $N(0)$,
increase as the square root of the Fermi level,
so that the number of electrons increases smoothly from zero.
This explains why a substantial number of carriers needs to be inserted in 3D semiconductors 
to achieve superconductivity \cite{Ekimov2004}. In a phonon-mediated
mechanism, T$_c$ is often proportional to the density of states at the Fermi level.
In 2D, as the density of states (DOS) is constant,
one would expect a constant $T_c$ as long as the phonon spectrum is weakly affected by doping.

This is in stark contrast with what happens in Li$_x$ZrNCl in the low
doping limit. This layered system can be considered the prototype of 2D 2-valley
electron gas. Indeed the bottom of the  conduction band of ZrNCl is
composed of  two perfectly parabolic bands at points ${\bf K}$ and
${\bf K^{\prime}}=2{\bf K}$ in the Brillouin zone. The interlayer
  interaction is extremely weak \cite{Heid2005,Takano2011,
    Botana2014, Paolo2015}.  Upon Li intercalation, the semiconducting
state is lost and superconductivity emerges. Surprisingly, the
superconducting critical temperature $T_c$ is strongly enhanced in the 
low-doping limit  \cite{Yamanaka1996,Yamanaka1998,Taguchi2006}, 
despite essentially parabolic bands, two-valley electronic structure
and an almost constant DOS \cite{Heid2005, Takano2011, Botana2014}.

In a 2D 2-valley electron gas,
the reduction of doping implies an increase of the $r_s$
electron-gas parameter 
and, consequently, of the electron-electron interaction \cite{GiulianiVignale}.
Then it can be expected that in the low doping limit the electronic structure, the vibrational properties,
and the electron-phonon interaction are strongly affected.
This is confirmed by the behavior of the magnetic
susceptibility. Despite Li$_x$ZrNCl being nonmagnetic, in the low-doping limit, 
the magnetic susceptibility is enhanced in a way very similar
way to that in the superconducting T$_c$. The interacting magnetic susceptibility 
$\chi_s$ is not constant, and strongly deviates from the constant
free-electron-like behavior in 2D \cite{Kasahara2009,Taguchi2010};
in particular, the susceptibility $\chi_s$ is enhanced at the low-doping regime.

In a previous work\cite{Paolo2015}, the behavior of $\chi_s$ and $T_c$ as a function of
doping was investigated
by using local functionals and a 2D two-valley electron gas model
solved within the random phase approximation (RPA). 
In this framework, it was found that the electron-electron interaction enhances the
electron-phonon matrix elements of those intervalley phonons inducing an unbalance in the valley occupations. 
The enhancement increases by increasing the $r_s$ parameter or, equivalently, 
by decreasing the electron density. 

In this paper, we perform a systematic study of electronic, magnetic, and vibrational properties
of Li$_x$ZrNCl using density functional theory (DFT) beyond the
standard LDA/GGA
approximations. We investigate the effect of an exact exchange
component on these properties and discuss the relevance of
electron-electron interaction in determining the superconducting
properties of Li$_x$ZrNCl.

In the following section, we show the structures used in our calculations.
In Sec. III, we present the technical details of our calculations.
In Sec. IV, we present the results of the electronic structure,
magnetic properties of valley and spin susceptibility,
phonon frequencies, and
electron-phonon coupling.
In the final section, we conclude our work.

\section{Crystal Structure}

\begin{figure}[!ht]
        \centering
        \includegraphics[clip=true, trim=0mm 0mm -10mm 0mm,width=0.5\textwidth]{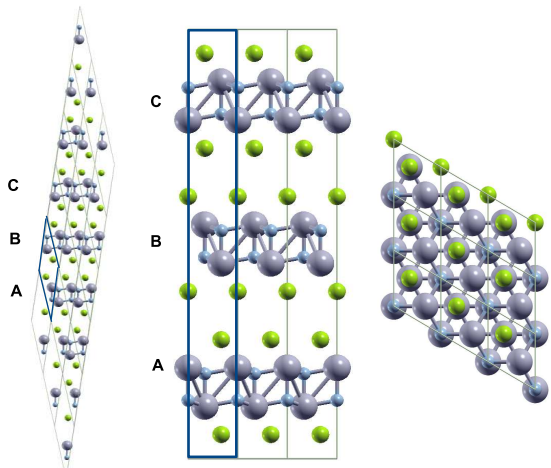}%left bottom right top
        \caption{ZrNCl structure. 
        Left: Rhombohedral cell repeated $3a \times 3b \times 3c$ to show the stacking.
        The unit cell with 2 formula units is highlighted.
        Middle: Hexagonal conventional cell repeated $3a \times 3a \times c$ to show the stacking with the side view along $\hat{a}-\hat{c}$ plane. 
        The conventional unit cell with 6 formula units is highlighted.
        Right: Top view along $\hat{a}-\hat{b}$ plane.}
        \label{fig:struc}
\end{figure}

The structures of undoped $\beta$-ZrNCl and Li-doped ZrNCl have been 
investigated using synchrotron x-ray \cite{Chen2001,Taguchi2006}
and powder neutron diffraction \cite{Shamoto1998}.
The primitive unit cell of ZrNCl has rhombohedral structure (space
group $R\bar{3}m$, number $166$) with 2 formula units per unit cell.
It can be also be  constructed by a conventional cell of hexagonal structure 
with 6 formula units per cell,
as shown in Fig. \ref{fig:struc}, where the ABC
layer stacking is evident.

Upon Li intercalation, Li atoms are placed between the ZrNCl layers.
Li acts as a donor and gives electrons to the Zr-N layers.
It has been shown that the Li intercalation can be simulated by
including an effective background charge, both for what concerns the
electronic structure and the phonon dispersion \cite{Heid2005,Botana2014}. %Weht1999,
Thus, we simulate Li doping by changing the number of electrons and 
using a compensating jellium background.
In all our calculations, the lattice parameters $a$ and $c$
are fixed to experimental values of each doping
\cite{Shamoto1998,Chen2001,Taguchi2006}
and the atomic coordinates are relaxed within a fixed volume.

\begin{figure}[!ht]
        \centering
                \includegraphics[clip=true, trim=0mm 0mm 0mm 0mm, width=0.35\textwidth]{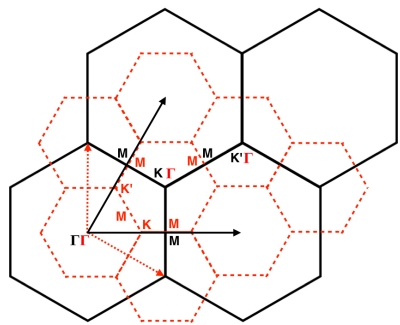}
        \caption{Brillouin zone of the hexagonal lattice of ZrNCl.
        		Large solid black hexagons represent the unit cell and
        		small dashed red hexagons represent the $\sqrt{3} \times \sqrt{3} \times 1$ supercell.
        		Note that the \textbf{K} and \textbf{K'} of the unitcell
        		fold onto the $\Gamma$ point of the supercell.}
        \label{fig:BZ}
\end{figure}

In order to be able to carry out the finite-difference
electron-phonon coupling calculation at the special point ${\bf K}$,
we take advantage of the weak interaction between the layers \cite{Kasahara2010},
making the stacking order negligible.
Therefore, we adopt the ZrNCl structure
with the lattice parameter $a$ set to the experimental value of each doping
and  simulate a single layer by inserting 12.5 \AA ~vacuum
between one ZrNCl layer and its periodic image corresponding to
$c=18.734$ \AA.
This is equivalent to the hexagonal structure with 
the space group $P\bar{3}m1$ (space group number 164),
with 2 formula units in the unit cell.
Then we create a supercell with the lattice vectors
$\sqrt{3}a\times\sqrt{3}a\times c$,
with 6 formula units.
In the Brillouin zone associated with the 
$\sqrt{3}\times\sqrt{3}\times1$ supercell of the hexagonal structure,
the special points ${\bf K}$ and ${\bf K^{\prime}}$ fold at ${\bf \Gamma}$, as shown in Fig. \ref{fig:BZ}.

\section{Computational Details}

Calculations are performed using the Hartree-Fock (HF) approximation and 
various flavors of DFT:
the Perdew-Zunger parametrization of the local density approximation (LDA) \cite{Dirac1930,PZ} ,
and generalized gradient approximation (GGA) as implemented in PBE \cite{PBE};
hybrid functionals with different exact exchange components, i. e., 
B3LYP \cite{hybrid1,blyp2,b3lyp} and
PBE0 ~\cite{PBE0};
and the range-separated HSE06 ~\cite{HSE06} hybrid functional.
The CRYSTAL14 periodic \textit{ab initio} code \cite{Crystal14} is used with 
norm-conserving pseudopotentials and 
Gaussian-type triple-$\zeta$ valence polarized basis sets \cite{TZVPbasis}
with the most diffuse Gaussian functions
of the Zr basis set reoptimized for periodic calculations. 
In order to check the accuracy of the Gaussian basis sets, the electronic band structure 
is compared to that obtained from a plane wave basis set calculation
performed using the {\sc Quantum ESPRESSO} method \cite{QE} with the PBE functional \cite{PBE}.

The doping of the semiconductor is simulated by changing the number of electrons on a compensating
jellium background, 
which has been previously shown to give accurate results for this system \cite{Heid2005,Botana2014}. 
The atomic coordinates are relaxed with lattice parameters fixed at the 
experimental values. 
For the energy convergence, a tolerance on the change in total energy of $10^{-9}$ Ha 
is used for all calculations.
A Fermi-Dirac smearing of 0.0025 Ha;
shrinking factors of 48-48, corresponding to an electron-momentum grid of $48 \times 48 \times 48$;
and real space integration tolerances of 7-7-7-15-30
are used for the relaxation of the internal coordinates 
and calculating the electronic band structure \cite{manualcrystal14}.
With this method, the exact exchange is computed in direct space, 
so the $k$ grids and $q$ grids of the electron momentum 
for the functionals with the exact exchange 
are equivalent.
The density of states is calculated using a Gaussian smearing of 0.005 Ha.

The effective mass, $m^*$ is calculated from the curvature of a fourth order polynomial fit to the 
region between the Fermi energy and the conduction band minimum around the special point,
\textbf{K}, assuming that the mass tensor is isotropic.
The fit parameters are given in Appendix \ref{app:meff}.

The spin susceptibility, $\chi_s$, obtained from 
the curvature of the energy as a function of total magnetization, $M$, is more sensitive
to the smearing and the $k$ grid.
Therefore, a smaller Fermi-Dirac smearing temperature of 0.00125 Ha
and finer shrinking factors of 120-120
are used to obtain energy as a function of magnetization, $E(M)$.

Electron-phonon coupling matrix elements and phonon frequencies are calculated with 
Fermi-Dirac smearing of 0.0035 Ha
and shrinking factors of $24 \times 24 \times 1$ in the $\sqrt{3}\times\sqrt{3}\times1$ supercell of
a single 2D lattice with AAA stacking.
The bands with the HF approximation in Sec. \ref{sec:HFsuscep} are also plotted
with these parameters.
For the electron-phonon coupling calculations,
the atoms are displaced according to the phonon pattern of the mode.
As this pattern is determined only by symmetry,
we use the same pattern as a function of doping.

\section{Results}

\subsection{Electronic Structure}

Undoped $\beta$-ZrNCl is a large gap insulator.
The direct band gap of the insulating compound is measured to be 3.4 eV with optical absorption spectra \cite{Ohashi1989}, 
while the indirect band gap of the Na$_{0.42}$ZrNCl is measured to be 2.5 eV with valence-band photoemission \cite{Yokoya2004}.
When doped, Li intercalation acts as a rigid filling of the parabolic
conduction band minima (valleys) at \textbf{K} and $\mathbf{K'}=2\mathbf{K}$ of the Brillouin zone,
leading to two quasispherical Fermi surfaces. 
We have calculated the electronic band structure with different levels of approximations
to evaluate the effect of the exact exchange on the electronic structure, band gap, and
effective mass.

\begin{figure}[!ht]
        \centering
                \includegraphics[clip=true, trim=-5mm 0mm 0mm -25mm, width=0.45\textwidth]{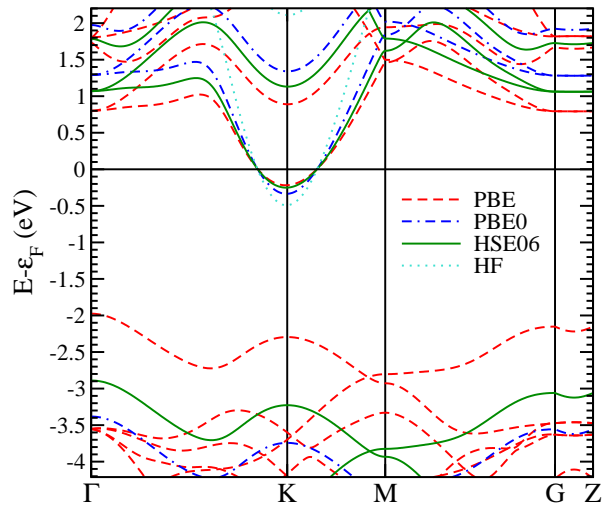}%left bottom right top
        \caption{Electronic structure of Li$_x$ZrNCl at doping $x=1/18$ with different functionals. 
        The Fermi level is set to 0 eV. 
        HF stands for Hartree-Fock.}
        \label{fig:bands}
\end{figure}

\begin{figure}[!ht]
        \centering
                \includegraphics[clip=true, trim=-5mm 0mm 0mm -25mm, width=0.45\textwidth]{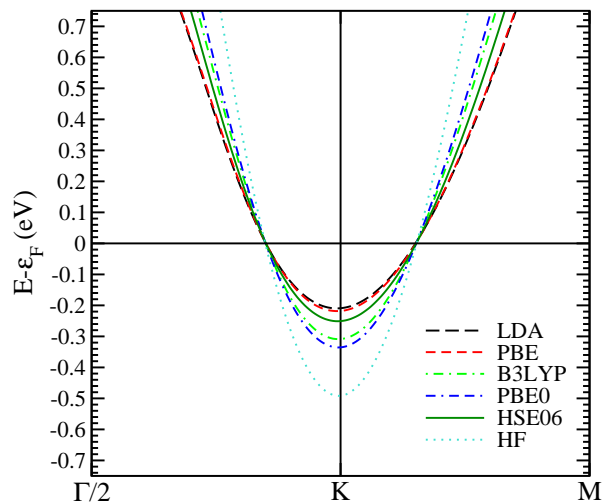}%left bottom right top
        \caption{A detailed view of the electronic structure of
          Li$_x$ZrNCl at doping $x=1/18$ around the conduction band minimum. 
        The Fermi level is set to 0 eV. }
        \label{fig:bandsEf}
\end{figure}

\begin{figure}[!ht]
        \centering
                \includegraphics[clip=true, trim=0mm 0mm -20mm -20mm, width=0.45\textwidth]{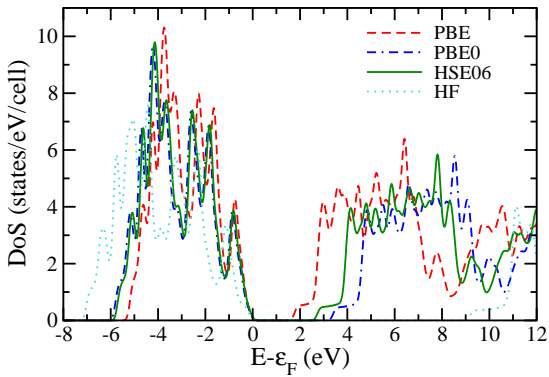}  %left bottom right top
        \caption{Density of states of undoped $\beta$-ZrNCl, 
        with Fermi level set to the top of the valence band. 
        The bottom of the conduction band has a quasiconstant density
      of states, a fingerprint of the 2D parabolic character of the electronic structure.}
        \label{fig:dos}

\end{figure}

\begin{figure}[!ht]
        \centering
                \includegraphics[clip=true, trim=0mm 0mm -10mm -10mm, width=0.45\textwidth]{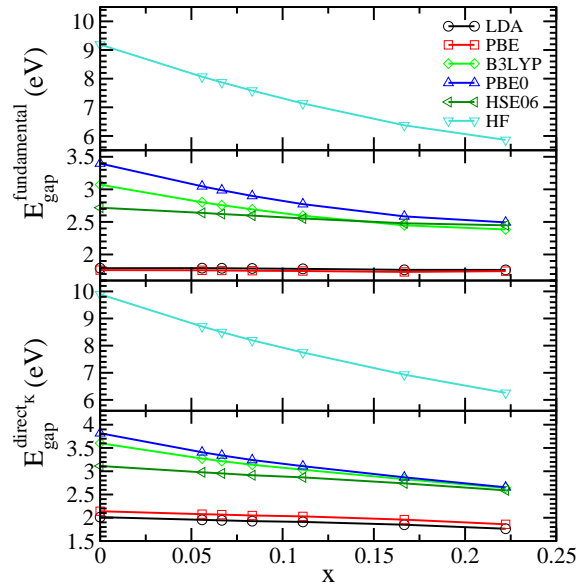}
        \caption{Change in the fundamental band gap with HF approximation (top panel) 
		        and DFT functionals (second panel),
    			and in the direct band gap at the $\mathbf{K}$ point with HF approximation (third panel) 
        		and DFT functionals (bottom panel) as a function of doping with different approximations.}
        \label{fig:Eg}
\end{figure}

The calculated electronic band structure is shown in Fig. \ref{fig:bands} and a detailed view around the conduction band minimum is shown in Fig. \ref{fig:bandsEf} for the lowest calculated doping of $x=1/18$.
Our PBE band structure is in agreement with previous calculations \cite{Botana2014}.
We also present the density of states of the undoped structure in Fig. \ref{fig:dos}.
The fundamental band gap is between the $\Gamma$-point and the $\mathbf{K}$-point.
We have also calculated the change in the direct band gap at the $\mathbf{K}$-point 
with different approximations.
These results are presented in Fig. \ref{fig:Eg}.

The LDA and PBE approximations produce similar results;
the electronic bands and the density of states are almost indistinguishable.
As a percentage of exact exchange is introduced with B3LYP and PBE0 functionals,
the band gap increases with increasing exchange fraction.
This becomes extreme in the HF limit, with a much larger band gap.
Therefore, there is a clear trend on how the electronic structure is modified with introduction
of the exact exchange in the approximations: 
the larger the exact exchange is, the larger is the calculated band gap.
For example, for the lowest doping of $x=1/18$, the calculated band gap changes as follows: 
with LDA and PBE $E_g \sim 1.8$ eV,
with B3LYP and PBE0 $E_g=2.8$ eV and $E_g=3.0$ eV respectively,
and finally with HF (which is the most extreme case) $E_g=8.1$ eV.
However, the introduction of the range separation together with the exact exchange
breaks this pattern.
With the HSE06 functional, we obtain a gap that is in between the PBE and B3LYP results.
For  the lowest doping of $x=1/18$, the calculated band gap becomes $E_g=2.6$ eV,
and the HSE06 results are in very good agreement with the valence-band photoemission
measurement of the indirect band gap of $E_g=2.5$ eV.

Furthermore, how the band gap changes with increasing doping is different with different approximations.
The band gap essentially does not change in LDA and PBE approximations,
with a decrease of $<0.02$ eV between the lowest and the highest doping.
However, in the same doping range, the gap decreases by 0.4 eV for B3LYP and 0.5 eV for PBE0.
The most extreme difference of $\sim$ 2 eV is obtained again with the HF approximation.
Hence, the gap does not stay constant if the exact exchange is introduced.
When the range separation is introduced, the band gap still decreases with the increasing
doping, but the difference is in between the PBE and B3LYP functionals.
With the HSE06 functional, the gap decreases by $\sim$ 0.2 eV between the lowest and highest doping,
keeping the results in good agreement with the experimentally reported value.

\begin{figure}[!ht]
        \centering
                \includegraphics[clip=true, trim=-3mm 0mm 0mm -20mm, width=0.45\textwidth]{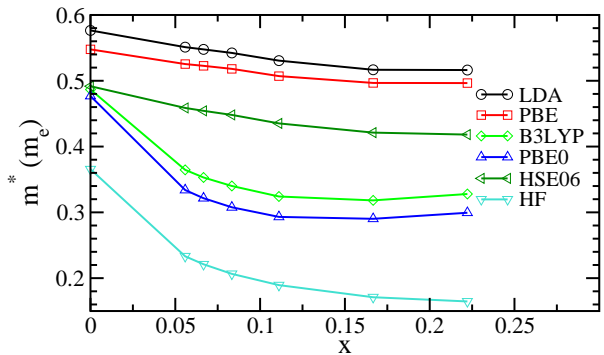}
        \caption{Change in the effective mass, $m^*$ as a function of doping with different approximations.}
        \label{fig:mass}
\end{figure}

Another result we can deduce from the electronic band structure is the change in the 
curvature of the conduction band, 
from which the effective mass, $m^*$, is calculated.
At fixed doping, as the exact exchange is introduced, the curvature of the conduction band gets larger,
leading to smaller effective mass.
This is also apparent in Fig. \ref{fig:bandsEf},
and can be seen when the PBE functional is compared to the hybrid B3LYP functional, and further to the PBE0 functional.
Intermediate steps of the B3LYP functional with the percentage of exact exchange changed to 5$\%$ and 10$\%$ can be found in Appendix \ref{app:exx}, and a gradual change in the effective mass and the band gap is observed.
The next step is introducing the range separation using the HSE functional family.
When the range separation of this functional is set to zero, i.e., $\omega=0$ Bohr$^{-1}$,
the PBE0 functional is recovered.
As the range separation parameter is increased to an intermediate value of 
$\omega=0.055$ Bohr$^{-1}$, the curvature starts to get smaller again,
resulting in increasing the effective mass. This is also presented in Appendix \ref{app:exx}.
The HSE06 functional with range separation $\omega=0.11$ Bohr$^{-1}$ further increases the effective mass,
giving a value between the PBE and B3LYP functionals.
The exact results are in Table \ref{table:param}.
The experimental value of the effective mass is 0.9 $m_e$ \cite{Takano2011};
however, this is an indirect derivation of the effective mass 
obtained from the optical reflectivity spectra using the Drude model, and
its value deviates from our HSE06 calculations.

The change in the $m^*$ as a function of doping is shown in Fig. \ref{fig:mass}.
While the change in the effective mass as a function of doping is almost constant for
PBE and LDA functionals, the introduction of the exact exchange with the B3LYP and PBE0 functionals
shows a difference of $\sim 0.15$ m$_{\rm e}$ between the undoped $x=0$
  and the lowest doped $x=1/18$ cases.
This difference is more dramatic with the HF approximation: the effective mass is too small
even in the undoped case and goes further down upon doping.
Finally, the HSE06 functional displays only a moderate decrease with increasing doping.

\subsection{Spin and Valley Magnetic Fields and Instabilities with the HF Approximation}
\label{sec:HFsuscep}
As shown in the previous section, the electronic structure of
Li$_x$ZrNCl is composed by two parabolic bands (valleys) at points ${\bf K}$ and
${\bf K^{\prime}}$ in the Brillouin zone. 
By adopting AAA stacking and
an in-plane $\sqrt{3}\times\sqrt{3}\times1$ supercell, 
the special ${\bf K}$ and ${\bf K^{\prime}}$ of the unit-cell Brillouin-zone 
fold at ${\bf \Gamma}$ in the Brillouin zone of the supercell. 
As a result, there will be
two perfectly degenerate parabolic bands at 
${\bf \Gamma}$ in the supercell Brillouin zone. 
By including the spin degrees of freedom, the total 
degeneracy at ${\bf \Gamma}$ of the supercell Brillouin zone is $4$.

\begin{figure}[h]
        \centering
                \includegraphics[clip=true, trim=0mm 0mm -10mm 0mm, width=0.5\textwidth]{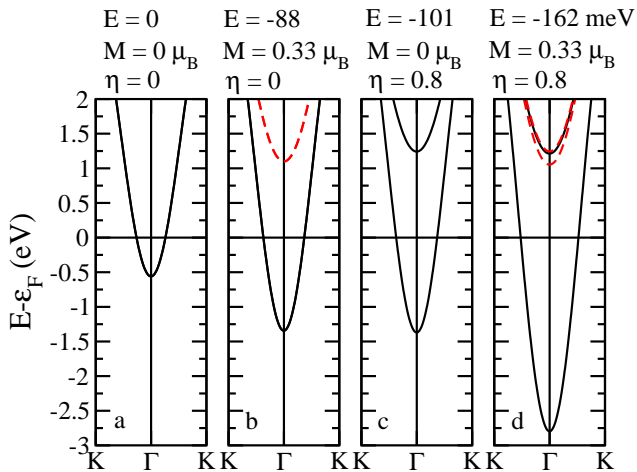}  %left bottom right top
        \caption{Electronic bands with HF approximation with $\sqrt3\times\sqrt3\times1$ cell at doping $x=1/18$.
		(a) Undistorted phase, with the displacement prefactor, $\eta=0$ [see Eq. (\ref{eq:eta})]; no magnetization $M=0$.
		(b) Undistorted phase under magnetization.
		(c) Distorted phase, no magnetization.
		(d) Distorted phase under magnetization. 
		The bands in (b) and (c) are obtained 
		at the energy minimum of the $E(M)$ and $E(\eta)$ curves, respectively.
		The difference between the energy of the structure in (a) and other structures
		is given in meV/cell (6 formula units) above each figure.
		The splitting between the two topmost bands in (d) is $\sim$ 0.03 eV.
		Red dashed lines represent the minority spin 
		and black solid lines represent the majority spin.
		The Fermi levels are shown by the black solid horizontal lines.}
        \label{fig:bandsHF}
\end{figure}

The nonmagnetic electronic structure calculated with the HF approximation and 
plotted in the $\sqrt{3}\times\sqrt{3}\times1$
Brillouin zone is shown in Fig. \ref{fig:bandsHF} (a).
If a spin unbalance is allowed (i. e., a finite magnetization), then
the spin degeneracy is broken and each one of the two bands splits in
two twofold-degenerate bands [see Fig. \ref{fig:bandsHF} (b)]. In the
Brillouin zone of the unit cell, this would mean that the spin
degeneracy in the valleys at ${\bf K}$ and ${\bf K^{\prime}}$ is broken
in the same way, as these two special points are still equivalent 
due to the unbroken $C_6$ symmetry.

Within the HF approximation, as a finite magnetization is introduced to the system, the energy goes down,
signifying that the HF approximation favors the magnetic state as the ground state.
For $x=1/18$, the energy of
the undistorted system under magnetization in Fig. \ref{fig:bandsHF} (b) is 88 meV/cell (6 formula units) lower than
the undistorted and nonmagnetic system in Fig. \ref{fig:bandsHF} (a).
This region of instability with the HF approximation continues up to the doping $x=1/6$,
as will be discussed in the following section.

It has been shown in Ref. \onlinecite{Paolo2015} that all phonons strongly
coupled to electrons and having phonon momentum 
${\bf q}={\bf K},{\bf K^{\prime}}$ (intervalley phonons) 
act as pseudo-magnetic fields, namely 
induce an asymmetry in the valley occupation, without breaking the
spin degeneracy, at least as long as there is no net magnetization.
Thus the intervalley distortion shifts the two valleys, changes the
occupation per valley but preserves the absence of a magnetization in
each valley. The action of an intervalley phonon on the electronic
structure at zero magnetization is shown in Fig. \ref{fig:bandsHF} (c).

Within the HF approximation, as the atoms are distorted along a phonon mode, the energy goes down;
signifying that the HF approximation favors the charge density state as the ground state.
For $x=1/18$, the energy of
the distorted and nonmagnetic system in Fig. \ref{fig:bandsHF} (c) is 101 meV/cell (6 formula units) lower than
the undistorted and nonmagnetic system in Fig. \ref{fig:bandsHF} (a).

Finally, Fig. \ref{fig:bandsHF} (d) shows the combined effect of
an intervalley distortion and a finite magnetization. The fourfold
degeneracy at ${\bf \Gamma}$ in the Brillouin zone of the supercell
is completely broken and $4$ different bands appear with different
spin and electron occupations. 

As expected, with the HF approximation, the distorted magnetic state has a lower energy
than the undistorted non-magnetic state.
For $x=1/18$, the energy of
the distorted and magnetic system in Fig. \ref{fig:bandsHF} (d) is 162 meV/cell (6 formula units) lower than
the undistorted and nonmagnetic system in Fig. \ref{fig:bandsHF} (a).

\subsection{Spin Susceptibility}

Magnetic properties are described by the spin susceptibility,
which is the response of the spin magnetization to an applied magnetic
field, namely:
\begin{equation}
\chi_s=\left(\frac{\partial^2 E}{\partial M^2}\right)^{-1}
\label{ChiM}
\end{equation}
where $E$ and $M$ are the total energy and magnetization, respectively.

The non interacting spin susceptibility, $\chi_{0s}$, is obtained
by neglecting the electron-electron interaction of the conducting electrons.
For perfectly parabolic bands, the non interacting spin susceptibility
is doping independent and equal to
\begin{equation}
\chi_{0s}=\mu_s N(0)=\frac{g_v m^{*}}{\pi \hbar^2}
\end{equation}
where $\mu_s$ is the Bohr magneton, $g_v$ is the valley degeneracy
($2$ in our case), and $m^{*}$ the band effective mass.
We calculate $\chi_{0s}$ from the density of states of the undoped compound,
which is shown in Fig. \ref{fig:dos},
and by extrapolating the $N(0)$ of the desired doping.
Our calculations show that $\chi_{0s}$ is not enhanced at the low-doping limit.

Experimental measurements \cite{Kasahara2009,Taguchi2010}  carried out
on Li$_x$ZrNCl show that 
(i) the system is not magnetic and 
(ii) the spin susceptibility in Li$_x$ZrNCl is strongly doping dependent 
with a marked enhancement in the low-doping limit 
\cite{Kasahara2009,Taguchi2010}, 
which is different than the expected $\chi_{0s}$ behavior.
It is then natural to look for exchange and correlation effects in the susceptibility.

We calculate the spin susceptibility with local and hybrid functionals
by finite differences. Namely we calculate the total energy at fixed magnetization
and then use Eq. (\ref{ChiM}) to obtain $\chi_s$. 

\begin{figure}[!ht]
        \centering
                \includegraphics[clip=true, trim=0mm -5mm -20mm 0mm, width=0.5\textwidth]{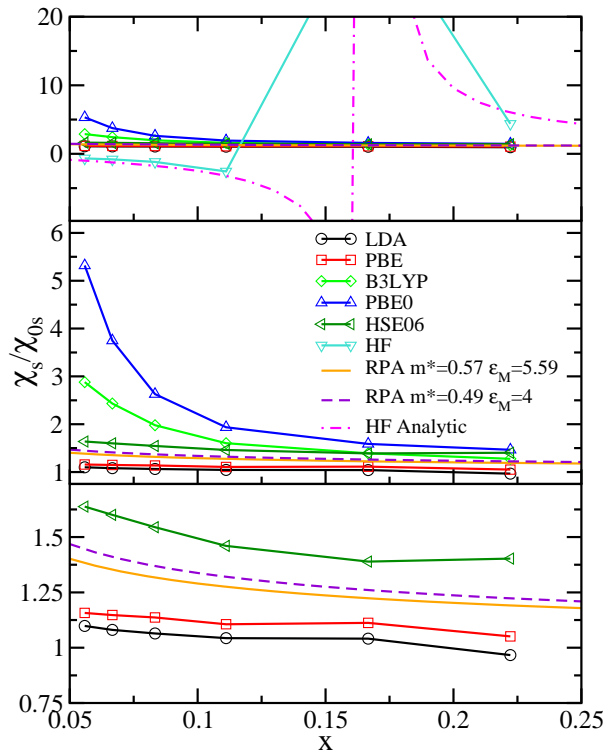} %left bottom right top
        \caption{Magnetic susceptibility enhancement factor, $\chi_s/\chi_{0s}$,
        obtained with different approximations.
        Each panel shows a detailed view of the panel above.}
        \label{fig:suscep}
\end{figure}

In Fig. \ref{fig:suscep}, we present the spin susceptibility enhancement factor $\chi_s/\chi_{0s}$ 
as a function of doping with different approximations.
The top panel of Fig. \ref{fig:suscep} displays the behavior of the enhancement factor with the HF approximation.
The HF approximation predicts that the nonmagnetic state is unstable in the low-doping limit.
As the magnetization is turned on, there is a finite gain in energy leading to a negative
spin susceptibility, $\chi_s$.
This result is in agreement with HF calculations carried out in
multivalley  2D electron gas \cite{Marchi2009}.
Furthermore, we have analytically calculated the spin susceptibility enhancement with the HF approximation,
considering the thickness of the 2D electron gas by including a form factor.
Details of the analytic expressions are given in Appendix \ref{section:HFAnalytic}.
The region of instability with the analytic HF calculation is similar 
to the numerical results,
proving that the form factor correctly takes into account the 
finite thickness of the 2D electron gas.

By reducing the amount of HF exchange in the functional, the spin
susceptibility enhancement at low doping is reduced, as shown in Fig. \ref{fig:suscep}. 
On the contrary, the bottom panel of Fig. \ref{fig:suscep} shows that
the LDA and PBE approximations show hardly any spin susceptibility
enhancement.
Thus, the susceptibility enhancement is entirely due to the
exchange interaction.
In Fig. \ref{fig:suscep}, we also compare our results 
with those obtained by a model based on RPA
\cite{DasSarma2005,Paolo2015} .
The model assumes a 2D 2-valley electron gas with no intervalley Coulomb scattering. 
Under this assumption, only the intravalley electron-electron
interaction remains and the RPA susceptibility can be calculated
analytically, by using the LDA/PBE effective mass of undoped ZrNCl
and the environmental dielectric constant, $\epsilon_M=5.59$ 
\cite{Takano2011,Botana2014,GalliPickett2010},
used in Ref. \onlinecite{Paolo2015}.
The model is appropriate in the low-doping limit where 
$|{\bf k_F}-{\bf K}|<<K$ (see Supplementary Material in Ref. \onlinecite{Paolo2015}),
a condition necessary to have the intravalley electron-electron scattering
dominating over the intervalley one.
 
As can be seen, the hybrid functional HSE06 gives an amount of enhancement 
from the high to low doping regime,
comparable to the one obtained with the RPA model.

\subsection{Phonon Frequencies}

In this section we evaluate the phonon frequencies of Li$_x$ZrNCl as a
function of doping for several functionals.

\begin{table*}[!hbt] \footnotesize
\caption{Frequencies corresponding to the two modes at the \textbf{K} point with high electron-phonon coupling,
		and six modes at the $\Gamma$ point which are Raman active and compared to the experimental values.
		All the frequencies are given in cm$^{-1}$.}
\centering
%\scalebox{0.5}{
%\begin{center}
\begin{ruledtabular}
	\begin{tabular}{l l c c c r c c c c} 
%	\hline
%	\hline
%	\cline{1-10}
	x & XC & K $\omega_1$ &	K $\omega_2$ & $\Gamma$ A$_{\rm 1g}$ & $\Gamma$ A$_{\rm 1g}$ & $\Gamma$ A$_{\rm 1g}$ & $\Gamma$ E$_{\rm g}$ & $\Gamma$ E$_{\rm g}$ & $\Gamma$ E$_{\rm g}$ \\
	\hline
        0    & Expt. [Ref. \onlinecite{Adelmann1999}] & $-$ & $-$ & 187 & 326 & 590 & 123 & 179 & 604 \\
        0.16 ($\sim$1/6)  & Expt. [Ref. \onlinecite{Adelmann1999}] & $-$ & $-$ & 188 & 322 & 582 & 123 & 178 & 608 \\
        0    & Expt. [Ref. \onlinecite{Cros2003}]     & $-$ & $-$ & 191 & 331 & 591 & 128 & 184 & 605 \\
        0                 & Expt. [Ref. \onlinecite{Kitora2007}] & $-$ & $-$ & 198 & 336 & 600 & $-$ & 191 & 614 \\
        0.06 ($\sim$1/18) & Expt. [Ref. \onlinecite{Kitora2007}] & $-$ & $-$ & 198 & 336 & 601 & $-$ & 190 & 614 \\
        0.10 ($\sim$1/9)  & Expt. [Ref. \onlinecite{Kitora2007}] & $-$ & $-$ & 197 & 331 & 592 & $-$ & 185 & 620 \\
        0.14 ($\sim$1/6)  & Expt. [Ref. \onlinecite{Kitora2007}] & $-$ & $-$ & 197 & 324 & 583 & $-$ & 181 & 613 \\
        0.24 ($\sim$2/9)  & Expt. [Ref. \onlinecite{Kitora2007}] & $-$ & $-$ & 195 & 326 & 585 & $-$ & 181 & 613 \\
        0.31 ($\sim$1/3)  & Expt. [Ref. \onlinecite{Kitora2007}] & $-$ & $-$ & 190 & 323 & 577 & $-$ & 178 & 603 \\
        \hline       
        0    & LDA   & 595 & 252 & 187 & 336 & 591 & 128 & 191 & 580 \\
        1/18 & LDA   & 484 & 227 & 186 & 332 & 587 & 127 & 188 & 584 \\
        1/15 & LDA   & 483 & 225 & 186 & 331 & 586 & 127 & 188 & 585 \\
        1/12 & LDA   & 483 & 220 & 186 & 330 & 584 & 126 & 187 & 587 \\
        1/9  & LDA   & 485 & 215 & 186 & 328 & 581 & 126 & 187 & 589 \\
        1/6  & LDA   & 498 & 203 & 184 & 323 & 574 & 123 & 185 & 593 \\
        2/9  & LDA   & 505 & 195 & 182 & 316 & 564 & 118 & 182 & 597 \\
        \hline                              
        0    & PBE   & 589 & 243 & 176 & 322 & 568 & 120 & 176 & 573 \\
        1/18 & PBE   & 484 & 219 & 176 & 318 & 563 & 118 & 174 & 578 \\
        1/15 & PBE   & 482 & 216 & 176 & 317 & 563 & 118 & 174 & 579 \\
        1/12 & PBE   & 481 & 211 & 176 & 315 & 561 & 117 & 174 & 581 \\
        1/9  & PBE   & 484 & 205 & 175 & 313 & 557 & 116 & 173 & 583 \\
        1/6  & PBE   & 491 & 195 & 173 & 306 & 549 & 112 & 171 & 587 \\
        2/9  & PBE   & 496 & 187 & 172 & 299 & 539 & 107 & 170 & 591 \\
        \hline                              
        0    & B3LYP & 608 & 252 & 179 & 330 & 586 & 123 & 179 & 585 \\
        1/18 & B3LYP & 347 & 207 & 178 & 325 & 581 & 121 & 176 & 588 \\
        1/15 & B3LYP & 373 & 206 & 178 & 324 & 580 & 121 & 177 & 589 \\
        1/12 & B3LYP & 419 & 207 & 178 & 322 & 578 & 119 & 176 & 591 \\
        1/9  & B3LYP & 457 & 213 & 178 & 319 & 576 & 118 & 176 & 594 \\
        1/6  & B3LYP & 486 & 208 & 176 & 313 & 569 & 112 & 174 & 599 \\
        2/9  & B3LYP & 496 & 199 & 175 & 306 & 557 & 108 & 173 & 603 \\
        \hline                              
        0    & PBE0  & 612 & 253 & 187 & 341 & 602 & 129 & 187 & 589 \\
        1/18 & PBE0  & 287 & 211 & 186 & 337 & 599 & 127 & 185 & 590 \\
        1/15 & PBE0  & 305 & 209 & 186 & 335 & 597 & 125 & 184 & 591 \\
        1/12 & PBE0  & 368 & 207 & 185 & 333 & 595 & 126 & 185 & 593 \\
        1/9  & PBE0  & 427 & 204 & 185 & 331 & 591 & 123 & 185 & 596 \\
        1/6  & PBE0  & 472 & 199 & 183 & 325 & 583 & 120 & 183 & 601 \\
        2/9  & PBE0  & 492 & 193 & 181 & 318 & 574 & 114 & 182 & 606 \\
        \hline                              
        0    & HSE06 & 611 & 252 & 186 & 339 & 601 & 128 & 186 & 588 \\
        1/18 & HSE06 & 453 & 223 & 185 & 335 & 596 & 126 & 185 & 591 \\
        1/15 & HSE06 & 451 & 220 & 185 & 333 & 595 & 126 & 184 & 591 \\
        1/12 & HSE06 & 458 & 217 & 185 & 332 & 593 & 125 & 184 & 594 \\
        1/9  & HSE06 & 468 & 211 & 185 & 330 & 590 & 123 & 183 & 597 \\
        1/6  & HSE06 & 485 & 201 & 167 & 326 & 584 & 120 & 182 & 602 \\
        2/9  & HSE06 & 493 & 193 & 181 & 316 & 572 & 114 & 181 & 604 \\
%	\hline
%	\hline
	\end{tabular}
	\end{ruledtabular}
%    }
%\end{center}
\label{table:freq}
\end{table*}  

We first compare the Raman-active phonon modes at the $\Gamma$ point 
with the experimental values \cite{Adelmann1999, Cros2003, Kitora2007},
as given in Table \ref{table:freq}. We find that all the functionals
are able to reproduce the Raman-active phonon modes within $\sim 10-20$ cm$^{-1}$
of the experimental values.
The LDA performs better than the PBE and B3LYP in reproducing the phonon frequencies at 
the $\Gamma$ point,
but the introduction of exact exchange into the PBE functional improves the results 
at the PBE0 and HSE06 levels
of approximation.

\begin{figure}[!ht]
        \centering
        \includegraphics[clip=true, trim=0mm -5mm 0mm 0mm, width=0.5\textwidth]{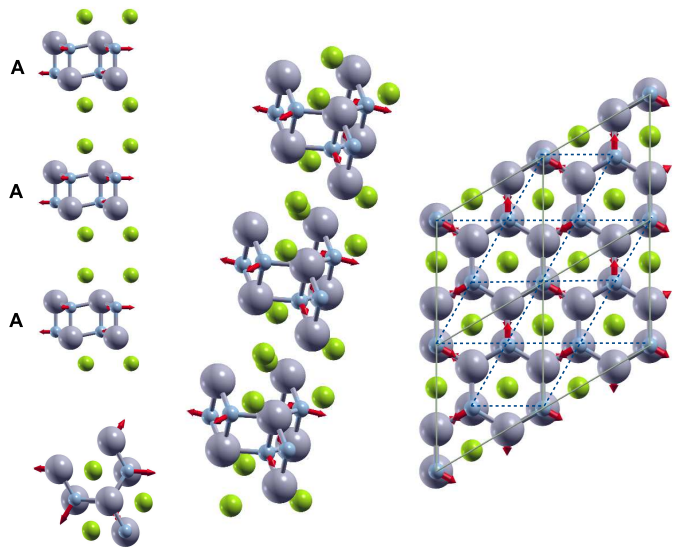}
        \caption{The phonon mode for $\omega_1$ at \textbf{K} point 
        shown in a $\sqrt{3}\times\sqrt{3}\times3$ supercell. 
        The structure is repeated along the $\hat{c}$ axis to show the AAA stacking
        as compared to the ABC stacking of Fig. \ref{fig:struc}.
        Left top: Side view along $\hat{a}-\hat{c}$ plane. 
        Left bottom: Top view along $\hat{a}-\hat{b}$ plane.
        Middle: Side view tilted to present the displacements.
        Right: Top view of $2\sqrt{3}\times2\sqrt{3}$ cell showing the 
        periodicity of the supercell with the solid gray lines,
        and the dashed blue lines show the $1\times1$ unit cell.
        }
        \label{fig:mode44}
\end{figure}

\begin{figure}[!ht]
        \centering
                \includegraphics[clip=true, trim=0mm 0mm -5mm -15mm, width=0.45\textwidth]{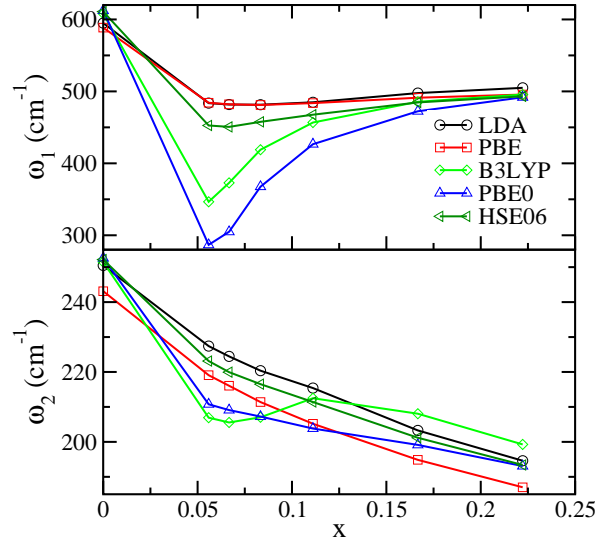}%left bottom right top
        \caption{Frequencies as a function of doping for the mode with high electron-phonon coupling
        with different approximations.}
        \label{fig:freq}
\end{figure}

Next, we calculate the phonon frequencies of intervalley phonons
(phonon momentum ${\bf q}={\bf K}$). 
In Ref.  \onlinecite{Paolo2015}, we establish that at the PBE level
the intervalley phonon with the highest electron-phonon
coupling has $\omega_1 \sim$ 59 meV (the associated phonon
displacement is shown in Fig. \ref{fig:mode44}). The second
mostly coupled intervalley phonon has $\omega_2 \sim 25$ meV. These
two modes account for the two main features in the Eliashberg function.
Thus we investigate in detail these two modes as a function of doping
and as a function of the exact exchange fraction.
The phonon frequencies  are  presented in Table \ref{table:freq}, and 
the behavior of frequency as a function of doping is shown in Fig. \ref{fig:freq}.

The intervalley phonon $\omega_1$ is softened significantly when doped, for all functionals.
This softening in the low-doping limit is weaker for the local functionals (LDA/PBE);
the change between the undoped and weakly doped modes is $\sim 100$ cm$^{-1}$.
It becomes substantial as the exact exchange fraction is enhanced, $\sim 300$ cm$^{-1}$ for PBE0.
Furthermore, the softening decreases as a function of doping.
In the case of PBE0 the softening at $x=1/18$ is $40\%$ of the phonon frequency at $2/9$.
It is worthwhile to stress that in the HF approximation (not shown here), 
as the non-magnetic state is unstable towards
a magnetic instability, (see Sec. \ref{sec:HFsuscep}), the phonon frequencies
are imaginary. We explicitly verify this by calculating the phonon
frequencies of the undistorted structure.

For the other intervalley phonon, $\omega_2$, 
the mode is also softened when doped,
but it decreases as a function of doping.
The softening of the frequency when doped is $\sim 30-50$ cm$^{-1}$,
and smaller than the softening of the $\omega_1$.
Therefore, the main contribution to the electron-phonon coupling 
comes from the phonon mode $\omega_1$.

In metals, a prominent softening of the phonon frequency is a
fingerprint of electron-phonon coupling. Thus the phonon frequency
calculations suggest that the intervalley electron-phonon coupling of
the mode $\omega_1$ is enhanced in the low-doping
limit in a way that is proportional to the amount of exact exchange present in
the functional, at least for what concerns non-range-separated
functionals.
The inclusion of range separation slightly decreases the phonon
softening, that, however, remains substantial in the low-doping limit.

\subsection{Electron-Phonon Coupling of Intervalley Phonons}

\begin{figure}[h]
        \centering
                \includegraphics[clip=true, trim=-5mm 0mm -5mm -16mm, width=0.45\textwidth]{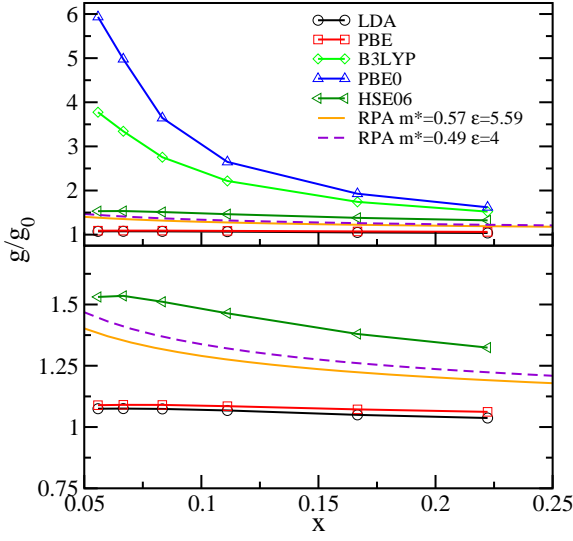}  %left bottom right top
        \caption{Electron-phonon coupling matrix elements
        ratio of the doped to the undoped system, $g/g_0$, 
        obtained with different approximations,
        shown in comparison with the $\chi_s/\chi_{0s}$ results obtained with the RPA calculations.
      	The panel below is a detailed view of the panel above.}
        \label{fig:elph}
\end{figure}

\begin{figure}[h]
        \centering
                \includegraphics[clip=true, trim=0mm 0mm -5mm -16mm, width=0.45\textwidth]{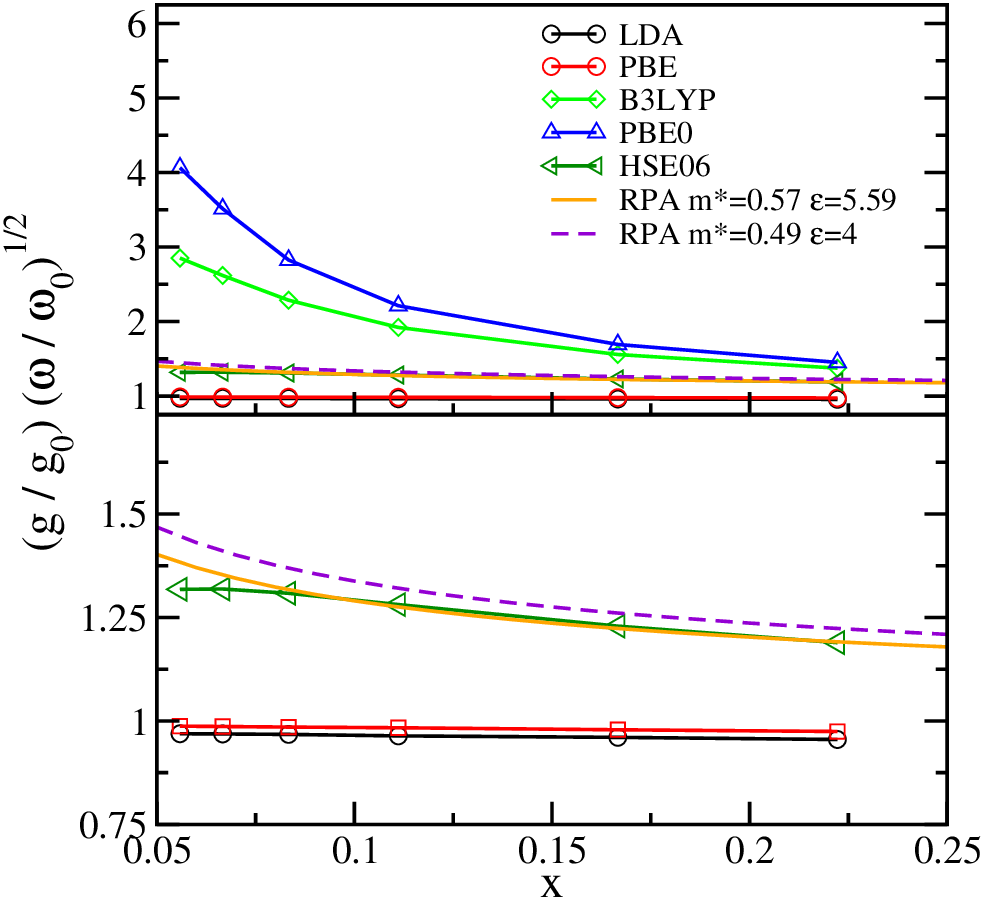}
        \caption{Electron-phonon coupling matrix elements
        ratio of the doped to the undoped system,
        scaled with the frequencies $\frac{g}{g_0}\sqrt{\frac{\omega}{\omega_0}}$,
        obtained with different approximations,
        shown in comparison with the $\chi_s/\chi_{0s}$ results obtained with the RPA calculations.
      	The panel below is a detailed view of the panel above.}
        \label{fig:elphrootw}
\end{figure}

The electron-phonon coupling matrix elements for a mode $\nu$ at a
phonon momentum $\mathbf{q}={\bf K}$ for electronic states
at the bottom of each valley, namely ${\bf k}={\bf K}$, are defined as
\begin{equation}
g^\nu_{\mathbf{K},\mathbf{2K}} = \sum_{A\alpha} \frac{e^{A \alpha}_{\mathbf{K}\nu}}{\sqrt{2M_A\omega_{\mathbf{K}\nu}}} \langle \mathbf{K} | \frac{\delta v_{SCF}}{\delta u^{\mathbf{K}}_{A\alpha}} | \mathbf{2K} \rangle ,
\label{eq:g2}
\end{equation}
where $A$ labels the atoms in the unit cell, $\alpha$ is the Cartesian coordinate,
and $u^\mathbf{K}_{A\alpha}$ is the Fourier transform of 
the phonon displacement of atom $A$ along direction $\alpha$,
with phonon frequency, $\omega_{\mathbf{K}\nu}$, and
$v_{SCF}$ is the periodic part of the screened potential.

The matrix element defined in Eq. (\ref{eq:g2}) can be calculated in a
frozen phonon approach. We consider a $\sqrt{3}\times\sqrt{3}\times1$
supercell. As both the special points ${\bf K}$ and ${\bf 2K}$ fold
at ${\bf \Gamma}$ when considering the supercell Brillouin zone, 
the electron-phonon matrix element in the supercell is
\begin{equation}
{\tilde g}^\nu_{\mathbf{\Gamma} n,\mathbf{\Gamma}m} =  \langle \mathbf{\Gamma}
n | \Delta V | \mathbf{\Gamma} m \rangle 
\label{eq:gtilde}
\end{equation}
where $m,n$ are band indexes running from $1$ to $2$. Indeed as the 
valleys at ${\bf K}$ and ${\bf 2K}$ in the Brillouin zone of the unit
cell now fold at ${\bf \Gamma}$ of the supercell, there are two
degenerate bands, each one twofold degenerate due to spin.
The operator $ \Delta V$ is defined as
\begin{equation}
\Delta V=\sum_{A\alpha} \frac{{\tilde e}^{A
    \alpha}_{\mathbf{\Gamma}\nu}}{\sqrt{2M_A\omega_{\mathbf{{\bf \Gamma}}\nu}}} \frac{\delta v_{scf}}{\delta u^{\mathbf{\Gamma}}_{A\alpha}}
\label{eq:dV}
\end{equation}
where now the Cartesian components of the phonon eigenvector ${\tilde
  e}^{A \alpha}_{\mathbf{{\bf \Gamma}}\nu}$ are normalized in the
$\sqrt{3}\times\sqrt{3}\times1$ supercell and can be chosen as real.

Equation \ref{eq:gtilde} can also be obtained in perturbation theory by
considering the Hamiltonian of the undistorted supercell $H_0$ and
$\Delta V$ as perturbation, namely,
\begin{equation}
H=H_0+\eta \Delta V
\label{eq:eta}
\end{equation}  
where $\eta$ is an arbitrary small constant that sets the magnitudes
of the perturbation or, equivalently, of the phonon displacement. 

The calculation of the electron-phonon matrix element in the supercell
amounts to calculating in first-order perturbation theory for
degenerate states the quantity $\langle n|\Delta V|m\rangle$.
The calculation can be simplified even more by noting that the
states $|\mathbf{\Gamma} m \rangle$ must be a linear combination of
the states  $|\mathbf{K} \rangle$ and $|\mathbf{K'} \rangle$ in the
Brillouin zone of the unit cell. As one can choose freely the states 
$|\mathbf{\Gamma} m \rangle$ in the degenerate subspace, we make
the choice  $|\mathbf{\Gamma} 1 \rangle=|\mathbf{K} \rangle$
and $|\mathbf{\Gamma} 2 \rangle=|\mathbf{2K} \rangle$. 
This choice assures that 
\begin{equation}
 \langle \mathbf{\Gamma}
n | \Delta V | \mathbf{\Gamma} n \rangle=0 
\end{equation}
as this matrix element couples electronic states at the same momentum
in the Brillouin zone of the unit cell, via a perturbation with a non-zero modulation.

So we are left with only the off-diagonal matrix
elements. By diagonalizing the matrix of the perturbation, we obtain that
the effect of the distortion on the electronic structure at linear
order is to split the two degenerate valleys at ${\bf \Gamma}$ (see
Fig. \ref{fig:bandsHF} (c) ) of an amount 
$\Delta \epsilon= 2\eta | \langle \mathbf{\Gamma}
1 | \Delta V | \mathbf{\Gamma} 2 \rangle|=2 \eta|{\tilde
  g}^\nu_{\mathbf{\Gamma} 1,\mathbf{\Gamma}2}|$.
Therefore we have the electron-phonon coupling in the supercell,
$|{\tilde
  g}^\nu_{\mathbf{\Gamma} 1,\mathbf{\Gamma}2}| = \frac{1}{2} \frac{d \Delta \epsilon}{d \eta} $,
that can be obtained by displacing the atoms in a way consistent with
the phonon displacement of the intervalley phonon and by performing
the derivative of the valley splitting as a function of the
distortion.

In order to relate the electron-phonon coupling of the supercell to
the one of the unit cell, we have to consider that the modes ${\tilde
  e}^{A \alpha}_{\mathbf{\Gamma}\nu}$
are normalized in the supercell, so that:
\begin{equation}
|g^\nu_{\mathbf{K},\mathbf{2K}}|^2 =3 |{\tilde
  g}^\nu_{\mathbf{\Gamma} 1,\mathbf{\Gamma}2}|^2
\end{equation}

For simplicity, in the rest of the discussion, we will denote
$g=|g^\nu_{{\bf K},{\bf 2K}}|$. 
To calculate the non-interacting electron-phonon coupling, $g_0$,
as in the case of the susceptibility, 
we use the insulating parent compound.
We then obtain the electron-phonon matrix elements by displacing the atoms along 
${\tilde e}^{A \alpha}_{\mathbf{\Gamma}\nu}$ 
obtained from a linear response run using the PBE functional 
and then by calculating the valley splitting in the supercell. 
As the relative magnitude of the different
Cartesian components in the phonon eigenvector are determined only by
the symmetry of the modes, the PBE eigenvector can then be used for
all the functionals, without introducing any error.

%%%%%%%%%%%%%%%%%%%%%%%%%%%%%%%%%%%%%%%%%%%%%%%%%%%%%%%%%%%%%%%%%%%%%%%%%%%%%%%%%%%%%%%%%%
\begin{table*}[!ht] \footnotesize
\caption{For each doping, and the functional, 
	the fundamental band gap E$_g$ between valence maximum at the $\Gamma$ point 
	and conduction minimum at the K point, 
	effective mass, $m^*$,
	density of states at the Fermi level, $N(0)$, 
	magnetic susceptibility, $\chi_s$, and 
	the enhancement factor, $\chi_s/\chi_{0s}$, of the magnetic susceptibility,
	electron-phonon coupling matrix element, $g$, 
	for the phonon mode $\omega_1$ at the $\mathbf{K}$ point,
	with $g_0$ calculated from the undoped $x=0$ case,
	and the enhancement factor, $g/g_0$, of the electron-phonon coupling matrix element.}
\centering
%\scalebox{0.5}{
%\begin{center}
	\begin{ruledtabular}
	\begin{tabular}{l l c c c d d c c} 
%	\hline
%	\hline
%	\cline{1-9}
	$x$ & XC & E$_g$ (eV) & $m^*$ (m$_e$) & N(0) (states/eV) & \multicolumn{1}{c}{$\chi_s$ ($\mu_{\rm B}^2$/eV)} & \multicolumn{1}{c}{$\chi_s/\chi_{0s}$} & $g$ (eV) & $g/g_{0}$ \\
	\hline
	0    & Expt. & 2.5 Ref.[\onlinecite{Yokoya2004}] & \\
	\hline
	0    & LDA   & 1.789 & 0.577 &                  & &  & 0.242 & 1.000 \\
	1/18 & LDA   & 1.790 & 0.551 & 0.535 & 0.588 & 1.099 & 0.260 & 1.074 \\
	1/15 & LDA   & 1.788 & 0.548 & 0.544 & 0.588 & 1.081 & 0.260 & 1.074 \\
	1/12 & LDA   & 1.785 & 0.542 & 0.552 & 0.588 & 1.065 & 0.260 & 1.074 \\
	1/9  & LDA   & 1.779 & 0.531 & 0.564 & 0.588 & 1.043 & 0.259 & 1.070 \\
	1/6  & LDA   & 1.764 & 0.517 & 0.598 & 0.623 & 1.042 & 0.254 & 1.050 \\
	2/9  & LDA   & 1.763 & 0.516 & 0.717 & 0.693 & 0.967 & 0.251 & 1.037 \\
	\hline
	0    & PBE   & 1.760 & 0.548 &                  & &  & 0.241 & 1.000 \\
	1/18 & PBE   & 1.757 & 0.525 & 0.521 & 0.603 & 1.157 & 0.263 & 1.091 \\
	1/15 & PBE   & 1.754 & 0.523 & 0.529 & 0.607 & 1.148 & 0.263 & 1.091 \\
	1/12 & PBE   & 1.750 & 0.518 & 0.539 & 0.613 & 1.137 & 0.263 & 1.091 \\
	1/9  & PBE   & 1.746 & 0.507 & 0.554 & 0.613 & 1.107 & 0.262 & 1.087 \\
	1/6  & PBE   & 1.732 & 0.497 & 0.595 & 0.662 & 1.113 & 0.259 & 1.075 \\
	2/9  & PBE   & 1.744 & 0.497 & 0.728 & 0.766 & 1.052 & 0.256 & 1.062 \\
	\hline  
	0    & B3LYP & 3.072 & 0.487 &                  & &  & 0.255 & 1.000 \\
	1/18 & B3LYP & 2.802 & 0.365 & 0.473 & 1.361 & 2.877 & 0.964 & 3.780 \\
	1/15 & B3LYP & 2.757 & 0.353 & 0.480 & 1.167 & 2.431 & 0.854 & 3.349 \\
	1/12 & B3LYP & 2.693 & 0.340 & 0.489 & 0.967 & 1.978 & 0.704 & 2.761 \\
	1/9  & B3LYP & 2.595 & 0.324 & 0.504 & 0.808 & 1.603 & 0.566 & 2.220 \\
	1/6  & B3LYP & 2.448 & 0.318 & 0.545 & 0.758 & 1.391 & 0.445 & 1.745 \\ 
	2/9  & B3LYP & 2.384 & 0.328 & 0.633 & 0.808 & 1.277 & 0.389 & 1.525 \\
	\hline
	0    & PBE0  & 3.397 & 0.477 &                  & &  & 0.266 & 1.000 \\
	1/18 & PBE0  & 3.047 & 0.334 & 0.461 & 2.450 & 5.315 & 1.577 & 5.929 \\
	1/15 & PBE0  & 2.987 & 0.322 & 0.467 & 1.750 & 3.747 & 1.322 & 4.970 \\
	1/12 & PBE0  & 2.899 & 0.308 & 0.474 & 1.246 & 2.629 & 0.968 & 3.639 \\
	1/9  & PBE0  & 2.773 & 0.293 & 0.487 & 0.942 & 1.934 & 0.704 & 2.647 \\
	1/6  & PBE0  & 2.585 & 0.290 & 0.520 & 0.826 & 1.589 & 0.512 & 1.925 \\
	2/9  & PBE0  & 2.493 & 0.299 & 0.570 & 0.835 & 1.465 & 0.430 & 1.617 \\
	\hline
	0    & HSE06 & 2.718 & 0.492 &                 &  &  & 0.261 & 1.000 \\
	1/18 & HSE06 & 2.639 & 0.459 & 0.472 & 0.774 & 1.640 & 0.400 & 1.533 \\
	1/15 & HSE06 & 2.621 & 0.454 & 0.478 & 0.766 & 1.603 & 0.401 & 1.536 \\
	1/12 & HSE06 & 2.597 & 0.448 & 0.486 & 0.750 & 1.543 & 0.395 & 1.513 \\
	1/9  & HSE06 & 2.553 & 0.435 & 0.498 & 0.728 & 1.462 & 0.382 & 1.464 \\
	1/6  & HSE06 & 2.479 & 0.421 & 0.532 & 0.739 & 1.389 & 0.360 & 1.379 \\
	2/9  & HSE06 & 2.448 & 0.418 & 0.582 & 0.817 & 1.404 & 0.346 & 1.326 \\
	\hline
	0    & HF    & 9.190 & 0.366 & \\
	1/18 & HF    & 8.068 & 0.233 & 0.356 & -0.237& -0.666 & \\
	1/15 & HF    & 7.871 & 0.221 & 0.360 & -0.291& -0.808 & \\
	1/12 & HF    & 7.584 & 0.207 & 0.366 & -0.430& -1.175 & \\
	1/9  & HF    & 7.134 & 0.189 & 0.376 & -0.967& -2.572 & \\
	1/6  & HF    & 6.373 & 0.171 & 0.405 & 14.700& 36.296 & \\
	2/9  & HF    & 5.860 & 0.165 & 0.441 &  1.934&  4.386 & \\
%	\hline
%	\hline
	\end{tabular}
	\end{ruledtabular}
%    }
%\end{center}
\label{table:param}
\end{table*}

The results of the calculation are shown in Fig. \ref{fig:elph} and
Table \ref{table:param}. 
While the electron-phonon matrix element is essentially constant when
using the LDA/PBE functionals, it is substantially enhanced in the low-doping
limit by the inclusion of exact exchange.
The enhancement in  the low doping limit decreases as the amount of
exact exchange decreases, 
namely in going from the PBE0 to  B3LYP functional, as shown in the top
panel of Fig. \ref{fig:elph}.
In HSE06, the enhancement is intermediate between PBE and B3LYP, due
to the introduction of range separation in the Coulomb term.
Table \ref{table:param} summarizes our results with the exact values of the E$_g$, $N(0)$, $\chi_s$, $\chi_s/\chi_{0s}$, $g$, and $g/g_0$ obtained up to this point.

There is a contribution of the softening in the phonon frequencies
to the enhancement of the electron-phonon matrix element,
as also evident from Eq. (\ref{eq:g2}).
To eliminate this contribution, we also plot, 
in Fig. \ref{fig:elphrootw},
the electron-phonon matrix elements
$\frac{g}{g_0}\sqrt{\frac{\omega}{\omega_0}}$,
where $\omega$ is the phonon frequency of the doped and $\omega_0$ is the phonon frequency of the undoped
structure.
Once the contribution of the phonon modes is removed, 
the agreement between the enhancement of the electron-phonon coupling of the HSE06 functional 
and the RPA calculation of $\chi_s/\chi_{0s}$ is improved.

We finally attempt to estimate the error due to the use of a localized
basis set on the electron-phonon coupling, by repeating our
calculation with the 
plane-wave basis sets within the {\sc Quantum ESPRESSO} method
using the PBE functional for the lowest doping $x=1/18$.
We find that the error in $g$ is $3.137 \%$ when using a localized basis set.

The reason for the enhancement of the electron-phonon matrix element
has been explained using a model RPA Hamiltonian in Ref. \onlinecite{Paolo2015}.
As shown in Ref. \onlinecite{Paolo2015} and in Fig. \ref{fig:bandsHF},
an intervalley phonon displacement can act as a pseudo-magnetic field by changing
the occupations of the valley at ${\bf K}$ and ${\bf K'}$ without invoking a finite
magnetization. In Ref. \onlinecite{Paolo2015} (Supplemental Material) it
was shown that as long as the intervalley Coulomb interaction can be
neglected, many-body effects enhance the valley susceptibility in the
same way as they enhance the spin susceptibility. Furthermore, it
was shown that the electron-phonon coupling of an intervalley phonon
inducing a valley polarization (pseudomagnetic field) should have an
enhancement electron-phonon interaction directly related to the
spin/valley susceptibility by the equation
\begin{equation}
\frac{g}{g_0}=\frac{\chi_s}{\chi_{0s}}
\label{eq:g_chi_enh}
\end{equation}

As the spin (and valley) susceptibility are strongly enhanced at low
doping by many-body effects, 
the same behavior should be found in the 
intervalley electron-phonon matrix element. 
Interestingly, following the work of Marchi {\it et al.} \cite{Marchi2009}, 
in a 2D 2-valley electron gas the spin susceptibility is mostly enhanced by the exchange interaction.
The source of divergence of $\chi_s/\chi_{0s}$ is the exchange interaction,
as the HF approximation is compared with the RPA calculation,
while the correlation effects,
taken into account with the Monte Carlo simulations in this work, 
bring this divergence down.
Because $r_s<1.5$ for Li$_x$ZrNCl, the RPA and Monte Carlo simulations
are identical in the regime of our interest.
Therefore, the main source of enhancement is the exchange interaction.
This result of enhancement due to the exchange interaction agrees with our findings. 
Indeed, the similar enhancement of
the electron-phonon interaction and of the spin susceptibility
confirms the validity of Eq. (\ref{eq:g_chi_enh}).

\section{Conclusion}

In this study, we have analyzed how the exchange and correlation affect the electronic, magnetic,
and vibrational properties of Li-doped ZrNCl, a two-dimensional two-valley semiconductor,
using different levels of approximations: HF, DFT with standard approximations, LDA and PBE, 
hybrid functionals with exact exchange B3LYP and PBE0,
and finally, a hybrid functional with exact exchange and range separation, HSE06.

By taking advantage of the parabolic conduction band minima,
we have calculated the change in the effective mass and band gap.
The HF approximation overestimates the band gap with respect to the experiments
and underestimates the effective mass, 
and similarly, the change in these properties as a function of doping is more drastic with
the hybrid functionals with exact exchange, PBE0 and B3LYP.
On the other hand, standard DFT approximations show almost constant band gap and effective mass
with changing doping.
The inclusion of the range separation provides a moderate change in these properties,
and the HSE06 results lie between the PBE and B3LYP functionals, 
and the band gap of the HSE06 functional is in good agreement with the experimental value.

The structure is unstable towards a magnetic and charge density state with the HF approximation.
Indeed, at low doping, as the magnetization is introduced, 
the HF approximation predicts the ground state of the system to be magnetic.
This presented itself as negative spin susceptibility up to the doping $x=1/6$, at which it diverges.
Parallel to this result, the larger the amount of the exact exchange in the hybrid functionals with PBE0 and B3LYP,
the larger is the spin susceptibility enhancement towards the low-doping regime.
On the other hand, the LDA and PBE approximations do not present any enhancement of the susceptibility.
Only HSE06, a hybrid functional with exact exchange and range separation, shows spin susceptibility
enhancement similar to the one obtained from the RPA calculations.

Next, the vibrational phonon modes and the electron-phonon coupling are calculated. 
The phonon frequency of the mode at the \textbf{K} point with high electron-phonon coupling
is softened significantly when a small doping is introduced to the system.
The frequency then increases as a function of doping,
and both the initial softening and the subsequent increase are larger, the larger the exact exchange.

We have calculated the electron-phonon coupling with the frozen phonon approach, 
by looking at the effect of the phonon displacement on the electronic bands.
%displays that the phonon is acting as a pseudo-magnetic field.
%
Analogously to how magnetic field lifts the spin degeneracy and splits the bands,
the phonon mode acts as a pseudo-magnetic field and lifts the valley degeneracy, splitting the electronic bands.
This is reflected in our results where
inter-valley electron-phonon matrix elements show a similar enhancement as compared to the 
enhancement in the spin susceptibility.

Therefore, we conclude that a phonon mode can act as a pseudomagnetic field and
electron-phonon interaction can cause an intervalley polarization.
The resulting electron-electron exchange interaction enhances the intervalley polarization, 
which in turn affects the superconducting temperature enhancement.
Furthermore, the differences between the standard density functionals
and those with exact exchange and range separation
imply that the description of the susceptibility
and electron-phonon coupling with a range-separated hybrid functional
would also be important in other 2D weakly doped semiconductors,
such as transition-metal dichalcogenides and graphene.

\begin{acknowledgments}
This work is supported by the Graphene Flagship and by Agence Nationale de la Recherche under
the reference no ANR-13-IS10-0003-01. Computer facilities were
provided by CINES, IDRIS and CEA TGCC (Grant EDARI No. 2016091202).
\end{acknowledgments}

\appendix
\section{Calculation of the Effective Mass}
\label{app:meff}

\begin{table}[!hbt] \footnotesize
\caption{The fit parameters to the $E(k)$.
		}
%	\centering
%\scalebox{0.5}{
\begin{center}
\begin{ruledtabular}
	\begin{tabular}{l l d d d} 
%	\hline
%	\hline
%	\cline{1-5}
	$x$ & XC & \multicolumn{1}{c}{$a_2 {\rm (eV/Bohr^{-2}})$} & \multicolumn{1}{c}{$a_3 {\rm (eV/Bohr^{-3}})$} & \multicolumn{1}{c}{$a_4 {\rm (eV/Bohr^{-4}})$} \\
	\hline
	0    & LDA   & 2.918 & -1.222 &  0.463 \\
	1/18 & LDA   & 3.008 & -1.098 &  0.123 \\
	1/15 & LDA   & 3.026 & -1.076 &  0.063 \\
	1/12 & LDA   & 3.056 & -1.041 & -0.061 \\
	1/9  & LDA   & 3.118 & -0.984 & -0.214 \\
	1/6  & LDA   & 3.251 & -0.856 & -0.542 \\
	2/9  & LDA   & 3.172 & -0.600 & -0.766 \\
	\hline              
	0    & PBE   & 3.071 & -1.232 &  0.089 \\
	1/18 & PBE   & 3.155 & -1.121 & -0.186 \\
	1/15 & PBE   & 3.171 & -1.097 & -0.234 \\
	1/12 & PBE   & 3.199 & -1.059 & -0.347 \\
	1/9  & PBE   & 3.263 & -0.979 & -0.534 \\
	1/6  & PBE   & 3.382 & -0.837 & -0.823 \\
	2/9  & PBE   & 3.299 & -0.577 & -1.024 \\
	\hline            
	0    & B3LYP & 3.453 & -1.164 & -0.387 \\
	1/18 & B3LYP & 4.548 & -1.084 & -2.244 \\
	1/15 & B3LYP & 4.695 & -1.064 & -2.425 \\
	1/12 & B3LYP & 4.877 & -1.035 & -2.607 \\
	1/9  & B3LYP & 5.102 & -0.983 & -2.766 \\
	1/6  & B3LYP & 5.281 & -0.894 & -2.519 \\ 
	2/9  & B3LYP & 4.993 & -0.693 & -2.052 \\
	\hline              
	0    & PBE0  & 5.398 &  0.486 & -0.244 \\
	1/18 & PBE0  & 4.964 & -1.070 & -2.812 \\
	1/15 & PBE0  & 5.153 & -1.056 & -3.049 \\
	1/12 & PBE0  & 5.387 & -1.044 & -3.287 \\
	1/9  & PBE0  & 5.643 & -1.008 & -3.346 \\
	1/6  & PBE0  & 5.787 & -0.955 & -2.792 \\
	2/9  & PBE0  & 5.470 & -0.779 & -2.158 \\
	\hline         
	0    & HSE06 & 3.421 & -1.196 & -0.110 \\
	1/18 & HSE06 & 3.614 & -1.106 & -0.449 \\
	1/15 & HSE06 & 3.648 & -1.088 & -0.495 \\
	1/12 & HSE06 & 3.699 & -1.058 & -0.576 \\
	1/9  & HSE06 & 3.801 & -0.998 & -0.781 \\
	1/6  & HSE06 & 3.979 & -0.908 & -0.979 \\
	2/9  & HSE06 & 3.916 & -0.658 & -1.158 \\
	\hline
	0    & HF    & 4.601 & -1.248 & -0.883 \\
	1/18 & HF    & 7.106 & -1.355 & -2.774 \\
	1/15 & HF    & 7.497 & -1.347 & -3.057 \\
	1/12 & HF    & 8.026 & -1.328 & -3.369 \\
	1/9  & HF    & 8.734 & -1.285 & -3.680 \\
	1/6  & HF    & 9.819 & -1.368 & -4.211 \\
	2/9  & HF    & 9.949 & -1.173 & -4.098 \\
%	\hline
%	\hline
	\end{tabular}
\end{ruledtabular}
%    }
\end{center}
\label{table:mass}
\end{table}    

We have calculated the effective mass by making a 
fourth-order polynomial fit to the conduction band,
along the direction of $\Gamma$ to K to M points of the Brillouin zone,
in a region around 0.1 eV above the Fermi level, and calculating the curvature at the band minimum.
In Table \ref{table:mass}, we present the fit parameters of the function: 
$E(k)=a_0 + a_1 (k-\mathbf{K}) + a_2 (k-\mathbf{K})^2 + a_3 (k-\mathbf{K})^3 + a_4 (k-\mathbf{K})^4$, 
with energy in units of eV and $k$ in units of $2\pi/a$.
As the absolute value of the energy is not known in the DFT framework,
we set the zero of the energy to the bottom of the conduction band,
making the constant term $a_0$ irrelevant.
The third- and fourth-order terms are important, 
because they show how much the Fermi surface
is warped with respect to that of the 2D electron gas.

The full expression for the dispersion $E(k)$
can be found in Ref. \onlinecite{Thomas2014}.
For simplicity, we have assumed that the anisotropy in the
effective mass tensor is small..
Hence, we have chosen the path $\Gamma$ to K to M
to take into account the conduction band minimum properly.
To understand the isotropy in the effective mass,
we calculated the effective mass with the same method
for $x=1/18$ with the PBE functional along the path
$\Gamma$ to K to $\Gamma$ and obtain $m^*=0.58$,
and M to K to M and obtain $m^*=0.50$,
as compared to the one obtained along the 
path of $\Gamma$ to K to M, $m^*=0.53$.

\section{Changing the Exact Exchange and Range Separation}
\label{app:exx}

In addition to the standard forms of the hybrid functionals, to understand the role of the
exact exchange percentage and the range separation, we have modified the parameters.

\begin{table}[!htb] \footnotesize
\caption{The band gap E$_g$ (eV), effective mass, m$^*$ (m$_e$), 
	density of states $N(0)$ (states/eV) , spin susceptibility $\chi_s$ ($\mu_{\rm B}^2$/eV),
	and the spin susceptibility enhancement factor, $\chi_s/\chi_{0s}$
	for the B3LYP functional with exact exchange percentage changed 
	to intermediate steps of $5\%$ and $10\%$,
	and for the HSE functional with the range separation parameter changed 
	to an intermediate value of $\omega=0.055$ \AA$^{-1}$
	and to $\omega=0.0$, which is the PBE0 limit.
		}
	\centering
%\scalebox{0.5}{
%\begin{center}
\begin{ruledtabular}
	\begin{tabular}{l c c c c c c} 
%	\hline
%	\hline
%	\cline{1-7}
	$x$ & XC & E$_g$ & m$^*$ & N(0) & $\chi_s$ & $\chi_s/\chi_{0s}$ \\
	\hline
	0    & $5\%$ & 2.079 & 0.527 &  & & \\
	1/18 & $5\%$ & 2.023 & 0.475 & 0.508 & 0.674 & 1.327 \\
	1/9  & $5\%$ & 1.969 & 0.446 & 0.545 & 0.639 & 1.173 \\
	1/6  & $5\%$ & 1.926 & 0.436 & 0.592 & 0.674 & 1.139 \\
	2/9  & $5\%$ & 1.917 & 0.439 & 0.789 & 0.774 & 0.981 \\
	\hline
	0    & $10\%$ & 2.400 & 0.512 &  & &  \\
	1/18 & $10\%$ & 2.279 & 0.434 & 0.495 & 0.808 & 1.632 \\
	1/9  & $10\%$ & 2.172 & 0.398 & 0.530 & 0.687 & 1.296 \\
	1/6  & $10\%$ & 2.094 & 0.389 & 0.574 & 0.700 & 1.220 \\
	2/9  & $10\%$ & 2.067 & 0.395 & 0.726 & 0.782 & 1.077 \\
	\hline
	0    & $\omega=0.0$ & 3.396 & 0.477 & & &  \\
	1/18 & $\omega=0.0$ & 3.047 & 0.333 & 0.460 & 2.450 & 5.315 \\
	1/9  & $\omega=0.0$ & 2.772 & 0.293 & 0.486 & 0.942 & 1.938 \\
	1/6  & $\omega=0.0$ & 2.582 & 0.290 & 0.520 & 0.826 & 1.589 \\
	2/9  & $\omega=0.0$ & 2.492 & 0.300 & 0.569 & 0.835 & 1.468 \\
	\hline
	0    & $\omega=0.055$ & 3.022 & 0.484 &  &  &  \\
	1/18 & $\omega=0.055$ & 2.857 & 0.421 & 0.466 & 1.097 & 2.354 \\
	1/9  & $\omega=0.055$ & 2.704 & 0.387 & 0.491 & 0.826 & 1.682       \\
	1/6  & $\omega=0.055$ & 2.580 & 0.372 & 0.525 & 0.790 & 1.505 \\
	2/9  & $\omega=0.055$ & 2.512 & 0.368 & 0.575 & 0.845 & 1.470 \\
	\end{tabular}
\end{ruledtabular}
%    }
%\end{center}
\label{table:exx}
\end{table}

We have changed the exact exchange percentage of the B3LYP functional from the original
$20\%$ to the intermediate values of $5\%$ and $10\%$.
In addition, if we set the range separation of the HSE functional to $\omega=0$, 
then the PBE0 functional must be recovered. 
We have performed the spin susceptibility calculations with this parameter to check this
limit of the HSE06 implementation.
Furthermore, we have changed the range separation to an intermediate value of $\omega=0.055$ \AA$^{-1}$,
to observe the change in our values as a function of the range separation parameter.
These results are presented in Table \ref{table:exx}.

We have observed that slowly increasing the amount of exact exchange in the hybrid functional
gradually changes the physical properties.
If the results of Table \ref{table:exx} are compared to those in Table \ref{table:param},
a gradual increase in the fundamental band gap, decrease in the effective mass,
and increase in the susceptibility are observed with increasing exact exchange percentage.

In addition, the HSE functional with the range separation set to $\omega=0$ gives the PBE0 limit
correctly, 
and the band gap, the effective mass, and the susceptibility are essentially the same.
Furthermore, changing the range separation of the HSE functional to an intermediate value of
$\omega=0.055$ \AA$^{-1}$, produces band gap, effective mass, and spin susceptibility 
values in between the original HSE06 and PBE0.
The divergence of the spin susceptibility in the low-doping limit with PBE0
decreases with increased range separation.
HSE06 functional gives reasonable results, but the perfect agreement with the experiments
would only be produced by a functional with adjusted exact exchange and range separation parameters.

\section{Spin Susceptibility as Compared to the Experiments}

\begin{figure}[!ht]
        \centering
              \includegraphics[clip=true, trim=0mm -5mm -20mm 0mm, width=0.5\textwidth]{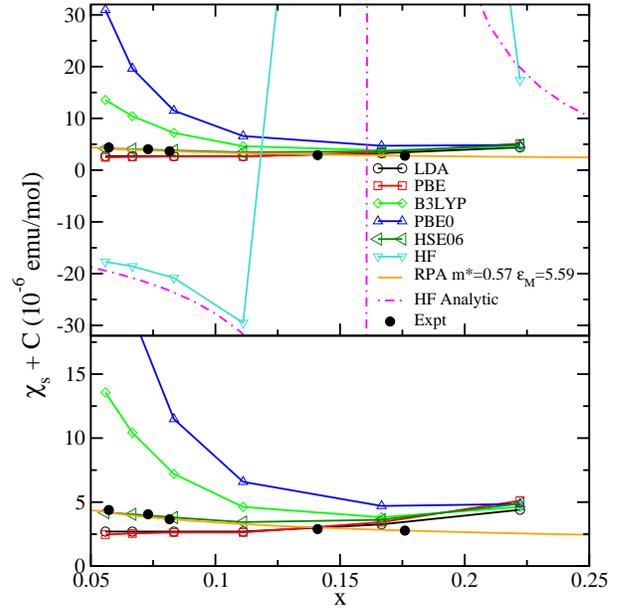}%left bottom right top
        \caption{Susceptibility $\chi_s$ obtained with different
        approximations as compared to the experimental data from
        Refs. \onlinecite{Kasahara2009,Paolo2015}.
        Panel below is a detailed view of the panel above.}
        \label{fig:chis}
\end{figure}

We present the interacting spin susceptibility $\chi_s$
as compared to the experiments in Fig. \ref{fig:chis}.
The experimental data are obtained from Ref. \onlinecite{Kasahara2009},
with the corrections as explained in
the Supplemental Material of Ref. \onlinecite{Kasahara2009} and
the Supplemental Material of Ref. \onlinecite{Paolo2015}.

The experimental data of spin susceptibility present the contributions
of all the electrons in the system, including the core electrons and
orbital contributions of the conducting electrons.
These terms are doping independent, and are subtracted from the experimental
data to obtain the spin susceptibility of the conduction electrons
in the experimental Ref. \onlinecite{Kasahara2009} and the corresponding 
Supplemental Material:
$\chi_s=\chi-\chi_{\rm core}^{\rm Li^+}-\chi_{\rm core}^{\rm ZrNCl}-\chi_L-\chi_{\rm orb}$.
The correct form of the Landau diamagnetic susceptibility is given in
Ref. \onlinecite{Paolo2015} and the corresponding Supplemental Material
as $\chi_L=-\chi_{0s}/(3m^{*2})$, and the experimental data presented
in Fig. \ref{fig:chis} show the corrected result.

\begin{table}[!htb] \footnotesize
\caption{The Landau diamagnetic susceptibility $\chi_L$ in $10^{-6}$emu/mol
		and the constant $C$ added to the calculated results
		for each exchange and correlation functional, XC,
		and HF approximation in both numerical and analytic calculation.
		RPA values are obtained from Ref. \onlinecite{Paolo2015}.
		}
	\centering
%\scalebox{0.5}{
%\begin{center}
\begin{ruledtabular}
	\begin{tabular}{l d d} 
%	\hline
%	\hline
%	\cline{1-7}
	XC & \multicolumn{1}{c}{$\chi_L$} & \multicolumn{1}{c}{$C$} \\
	\hline
	LDA   &  -8.81 &  -6.81 \\
	PBE   &  -9.27 &  -7.27 \\
	B3LYP & -10.43 &  -8.43 \\
	PBE0  & -10.65 &  -8.65 \\
	HSE06 & -10.32 &  -8.32 \\
	HF    & -13.89 & -13.89 \\
	RPA   &  -8.89 &  -7.77 \\
	\end{tabular}
\end{ruledtabular}
%    }
%\end{center}
\label{table:chiL}
\end{table}	
	
For the theoretical results in Fig. \ref{fig:chis},
we add a constant, $C$, to our calculations to account for the 
uncertainties in the spin susceptibility contribution of
other doping-independent terms.
To estimate this constant, we first calculate the
Landau susceptibility, $\chi_L$, for each approximation,
using the effective mass of the undoped structure.
A further shift is added to the HSE06 functional to match the
experimental enhancement of the susceptibility in the low-doping regime, 
and the same shift is added to all the other functionals for comparison.
Table \ref{table:chiL} shows the Landau susceptibility, $\chi_L$, and 
the constant shift, $C$, applied to the calculations in Fig \ref{fig:chis}.

\section{Spin and Valley Magnetic Fields with HSE06 functional}

\begin{figure}[h]
        \centering
                \includegraphics[clip=true, trim=0mm 0mm -10mm 0mm, width=0.5\textwidth]{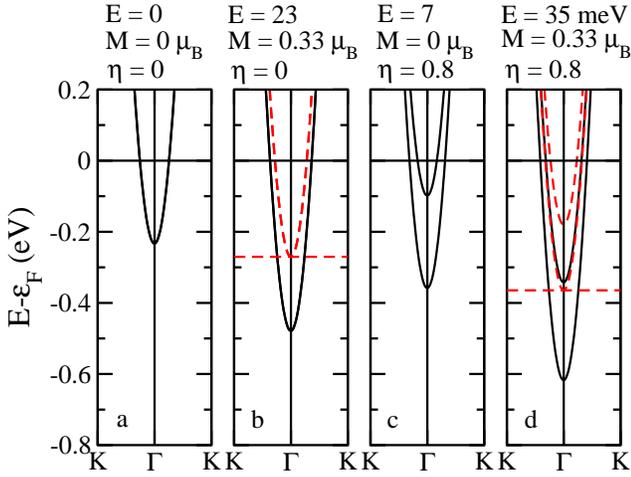}%left bottom right top
        \caption{Electronic bands with HSE06 functional with $\sqrt3\times\sqrt3\times1$ cell at doping $x=1/18$.
		(a) Undistorted phase, with the displacement prefactor, $\eta=0$ [see Eq. (\ref{eq:eta})]; no magnetization $M=0$.
		(b) Undistorted phase under magnetization.
		(c) Distorted phase, no magnetization.
		(d) Distorted phase under magnetization. 
		The difference between the energy of the structure in panel (a) and other structures
		is given in meV/cell (6 formula units) above each figure.
		Red dashed lines represent the minority spin 
		and black solid lines represent the majority spin.
		The Fermi levels of the nonmagnetic case and the majority spin are set to zero
		and are shown by the black solid horizontal lines.
		The Fermi level of the minority spin is shown by the red dashed horizontal line.
		}
        \label{fig:bandsHSE06}
\end{figure}

We present how the spin and valley degeneracy 
is lifted in the case of the HSE06 functional 
in Fig. \ref{fig:bandsHSE06}.
As the instabilities do not exist with this functional, 
we present the electronic bands
at the HF energy minimum of each instability 
for a comparison with the results
presented in Fig. \ref{fig:bandsHF}.

Figure \ref{fig:bandsHSE06} shows that the introduction of the magnetization and distortion
do not lead to an instability with the HSE06 functional,
because the energy always increases with respect to the undistorted, nonmagnetic system.
Since the system is not at the energy minimum in the case where the total magnetization
is fixed, as in Fig. \ref{fig:bandsHSE06} panels (b) and (d), 
the Fermi energy of the majority and minority spins are not the same, 
and the zero of the Fermi level of the majority spins is set to zero.
All the electrons of the $\sqrt{3}\times\sqrt{3}$ cell are polarized in this case.
The bands of the minority spin are not occupied; therefore the Fermi level
of the minority spin band is set to the minimum of the conduction band.

Furthermore, we also present the displacement of each atom for the mode $\omega_1$
corresponding to $\eta=0.8$ in Table \ref{table:displ0.8}.
This mode is dominated by the displacements of the N atoms along the $x$-$y$ direction,
and therefore is a breathing mode of the N atoms.

\begin{table}[!htb] \footnotesize
\caption{The displacements of each atom from the undistorted phase corresponding to $\eta=0.8$ for the mode $\omega_1$
	with large electron-phonon coupling at the {\bf K}-point.
	Distances are given in \AA.
                }
        \centering
%\scalebox{0.5}{
%\begin{center}
\begin{ruledtabular}
        \begin{tabular}{l d d d}
%       \hline
%       \hline
%       \cline{1-7}
        Atom & \multicolumn{1}{c}{$dx$} & \multicolumn{1}{c}{$dy$} & \multicolumn{1}{c}{$dz$} \\
        \hline
	Zr &   0.001042 & -0.000602 &  0.000000 \\
        Zr &  -0.001043 & -0.000602 &  0.000000 \\
        Zr &   0.000000 &  0.001204 &  0.000000 \\
        Zr &   0.000000 & -0.001060 &  0.000000 \\
        Zr &   0.000918 &  0.000530 &  0.000000 \\
        Zr &  -0.000918 &  0.000530 &  0.000000 \\
        N  &   0.013543 & -0.007819 &  0.000000 \\
        N  &  -0.013543 & -0.007818 &  0.000000 \\
        N  &  -0.000001 &  0.015637 &  0.000000 \\
        N  &   0.000000 & -0.017497 & -0.000002 \\
        N  &   0.015153 &  0.008748 & -0.000002 \\
        N  &  -0.015153 &  0.008749 &  0.000004 \\
        Cl &   0.000000 &  0.000000 &  0.000380 \\
        Cl &   0.000000 &  0.000000 & -0.000761 \\
        Cl &   0.000000 &  0.000000 &  0.000380 \\
        Cl &   0.000000 &  0.000000 & -0.000461 \\
        Cl &   0.000000 &  0.000000 &  0.000924 \\
        Cl &   0.000000 &  0.000000 & -0.000461 \\
        \end{tabular}
\end{ruledtabular}
%    }
%\end{center}
\label{table:displ0.8}
\end{table}

\section{Analytic HF Calculation of the Spin Susceptibility Enhancement}
\label{section:HFAnalytic}
The interacting spin susceptibility of multivalley 2D electron gas can be analytically calculated by
\begin{equation}
\frac{\chi_{0s}}{\chi_s} = 1- \frac{2 \alpha r_s}{\pi} \int_0^1 dx \frac{xF(x)}{\sqrt{1-x^2}}
\end{equation}
where $\alpha=\sqrt{g_v g_s/4}=1$ for a valley $g_v=2$ and spin $g_s=2$ degeneracy, and $q=2x k_F$.

The electron-gas parameter is defined as $r_s=1/(a_B\sqrt{\pi n})$,
where the electron density, $n$ is linked to the doping, $x$,
per area, $\Omega$, of 2 formula units of ZrNCl: $n=2x/\Omega$.
At the low-doping regime, Li$_x$ZrNCl has $r_s<1.5$.

\begin{figure}[!ht]
        \centering
              \includegraphics[clip=true, trim=0mm 0.7mm 0mm 0mm, width=0.45\textwidth]{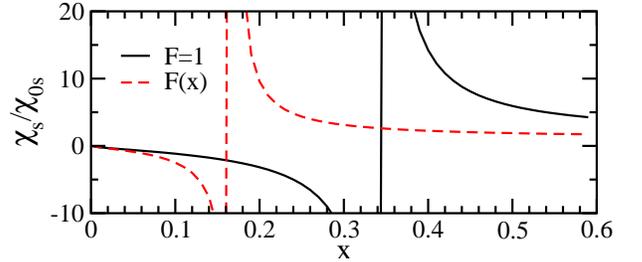}%left bottom right top
        \caption{The analytic calculation of the spin susceptibility enhancement with HF approximation,
        with, $F(x)$, and without, $F=1$, considering the thickness of the 2D electron gas.}
        \label{fig:form}
\end{figure}

For a strictly 2D electron gas, the form factor $F=1$ gives the textbook expression of the spin susceptibility 
enhancement within the HF approximation \cite{GiulianiVignale}:
$\chi_{0s}/\chi_s=1-2r_s/\pi$.

To take into account the thickness of the 2D electron gas, 
we obtain an additional
term to the form factor.
The form factor can be derived by considering the exchange energy
$E_x=v_qF(q)$, with the Coulomb potential, $v_q=2\pi/q$.
Then we first Fourier-transform 
the 2D Coulomb interaction with a certain component along the z-direction,
\begin{eqnarray}
FT\left[\frac{1}{r}\right] &=& 
\int d^2r_{\parallel} \frac{1}{\sqrt{r_{\parallel}^2+z^2}}
e^{-i\vec{q_{\parallel}}\cdot\vec{r_{\parallel}}} \\ \nonumber
&=& \frac{2\pi}{q} e^{-q|z|}
\end{eqnarray}
and further integrate along the z-direction for
an electron gas of a thickness of $a$,
\begin{eqnarray}
\frac{2\pi}{q}F(q) &=& \frac{2\pi}{q} \frac{1}{a^2} \int_{-a/2}^{a/2}dz_1
\int_{-a/2}^{a/2}dz_2 e^{-q|z_1-z_2|} \\ \nonumber
&=&\frac{2\pi}{q} \frac{2}{qa}\left( 1+ \frac{e^{-qa}-1}{qa} \right)
\end{eqnarray}
which leads to a form factor of
\begin{equation}
F(q)=\frac{2}{qa^*}\left( 1+ \frac{e^{-qa^*}-1}{qa^*} \right)
\end{equation}
with $q=2x k_F$ and
the thickness of the 2D electron gas is taken into account in
$a^*=a/a_B$ renormalized by $a_B=\epsilon_M \hbar^2 / (m^* e^2)$.
The thickness of the electron gas is estimated from 
the thickness of the ZrN bilayer along the $\hat{c}$ axis 
and is taken to be $a=2.5$ \AA~in our calculations.
We use the effective mass numerically obtained from the undoped HF structure and 
the environmental dielectric constant set to $\epsilon_M=1$.

In the analytic form of the Hartree-Fock approximation,
the metallic screening of the adjacent layers is not taken into account.
This is because the electron-electron interaction is not screened in the HF approximation.
On the other hand, with the RPA approximation, the electron-electron interaction is screened.
In addition, the hopping between the layers is negligible in the present case,
hence the HF exchange energy does not contain inter-layer contributions.
Indeed the HF exchange energy is equal to:
$(-1/2) \int d^3r d^3r' \rho({\bf r},{\bf r}')(1/|{\bf r}-{\bf r}'|)$,
where $\rho({\bf r},{\bf r}')$ are the off-diagonal elements of the density matrix.
The hopping between layers is negligible, $\rho({\bf r},{\bf r}')=0$,
if ${\bf r}$ and ${\bf r}'$ belong to different layers.

Figure \ref{fig:form} shows the difference between the spin susceptibility enhancement
with and without considering the thickness of the 2D electron gas.
When the thickness is not taken into account, i.e., $F=1$,
the compound is unstable in a larger region of doping, $x<0.34$,
while considering the thickness of the 2D electron gas
moves the instability region to $x<0.16$, 
which better agrees with the numerical calculations as shown in Fig. \ref{fig:suscep}.
The difference between the instability region with and without considering the thickness of the 2D electron gas, 
as well as the agreement between the analytic expression and the numerical calculations,
gives us confidence in using the form factor $F(x)$.

\bibliography{PaperBetul}

\begin{thebibliography}{47}%
\makeatletter
\providecommand \@ifxundefined [1]{%
 \@ifx{#1\undefined}
}%
\providecommand \@ifnum [1]{%
 \ifnum #1\expandafter \@firstoftwo
 \else \expandafter \@secondoftwo
 \fi
}%
\providecommand \@ifx [1]{%
 \ifx #1\expandafter \@firstoftwo
 \else \expandafter \@secondoftwo
 \fi
}%
\providecommand \natexlab [1]{#1}%
\providecommand \enquote  [1]{``#1''}%
\providecommand \bibnamefont  [1]{#1}%
\providecommand \bibfnamefont [1]{#1}%
\providecommand \citenamefont [1]{#1}%
\providecommand \href@noop [0]{\@secondoftwo}%
\providecommand \href [0]{\begingroup \@sanitize@url \@href}%
\providecommand \@href[1]{\@@startlink{#1}\@@href}%
\providecommand \@@href[1]{\endgroup#1\@@endlink}%
\providecommand \@sanitize@url [0]{\catcode `\\12\catcode `\$12\catcode
  `\&12\catcode `\#12\catcode `\^12\catcode `\_12\catcode `\%12\relax}%
\providecommand \@@startlink[1]{}%
\providecommand \@@endlink[0]{}%
\providecommand \url  [0]{\begingroup\@sanitize@url \@url }%
\providecommand \@url [1]{\endgroup\@href {#1}{\urlprefix }}%
\providecommand \urlprefix  [0]{URL }%
\providecommand \Eprint [0]{\href }%
\providecommand \doibase [0]{http://dx.doi.org/}%
\providecommand \selectlanguage [0]{\@gobble}%
\providecommand \bibinfo  [0]{\@secondoftwo}%
\providecommand \bibfield  [0]{\@secondoftwo}%
\providecommand \translation [1]{[#1]}%
\providecommand \BibitemOpen [0]{}%
\providecommand \bibitemStop [0]{}%
\providecommand \bibitemNoStop [0]{.\EOS\space}%
\providecommand \EOS [0]{\spacefactor3000\relax}%
\providecommand \BibitemShut  [1]{\csname bibitem#1\endcsname}%
\let\auto@bib@innerbib\@empty
%</preamble>
\bibitem [{\citenamefont {Saito}\ \emph {et~al.}(2015)\citenamefont {Saito},
  \citenamefont {Kasahara}, \citenamefont {Ye}, \citenamefont {Iwasa},\ and\
  \citenamefont {Nojima}}]{Saito2015}%
  \BibitemOpen
  \bibfield  {author} {\bibinfo {author} {\bibfnamefont {Y.}~\bibnamefont
  {Saito}}, \bibinfo {author} {\bibfnamefont {Y.}~\bibnamefont {Kasahara}},
  \bibinfo {author} {\bibfnamefont {J.}~\bibnamefont {Ye}}, \bibinfo {author}
  {\bibfnamefont {Y.}~\bibnamefont {Iwasa}}, \ and\ \bibinfo {author}
  {\bibfnamefont {T.}~\bibnamefont {Nojima}},\ }\href@noop {} {\bibfield
  {journal} {\bibinfo  {journal} {Science}\ }\textbf {\bibinfo {volume}
  {350}},\ \bibinfo {pages} {409} (\bibinfo {year} {2015})}\BibitemShut
  {NoStop}%
\bibitem [{\citenamefont {Lu}\ \emph {et~al.}(2015)\citenamefont {Lu},
  \citenamefont {Zheliuk}, \citenamefont {Leermakers}, \citenamefont {Yuan},
  \citenamefont {Zeitler}, \citenamefont {Law},\ and\ \citenamefont
  {Ye}}]{Ye2015}%
  \BibitemOpen
  \bibfield  {author} {\bibinfo {author} {\bibfnamefont {J.~M.}\ \bibnamefont
  {Lu}}, \bibinfo {author} {\bibfnamefont {O.}~\bibnamefont {Zheliuk}},
  \bibinfo {author} {\bibfnamefont {I.}~\bibnamefont {Leermakers}}, \bibinfo
  {author} {\bibfnamefont {N.~F.~Q.}\ \bibnamefont {Yuan}}, \bibinfo {author}
  {\bibfnamefont {U.}~\bibnamefont {Zeitler}}, \bibinfo {author} {\bibfnamefont
  {K.~T.}\ \bibnamefont {Law}}, \ and\ \bibinfo {author} {\bibfnamefont
  {J.~T.}\ \bibnamefont {Ye}},\ }\href@noop {} {\bibfield  {journal} {\bibinfo
  {journal} {Science}\ }\textbf {\bibinfo {volume} {350}},\ \bibinfo {pages}
  {1353} (\bibinfo {year} {2015})}\BibitemShut {NoStop}%
\bibitem [{\citenamefont {Novoselov}\ \emph {et~al.}(2005)\citenamefont
  {Novoselov}, \citenamefont {Jiang}, \citenamefont {Schedin}, \citenamefont
  {Booth}, \citenamefont {Khotkevich}, \citenamefont {Morozov},\ and\
  \citenamefont {Geim}}]{Novoselov2005}%
  \BibitemOpen
  \bibfield  {author} {\bibinfo {author} {\bibfnamefont {K.~S.}\ \bibnamefont
  {Novoselov}}, \bibinfo {author} {\bibfnamefont {D.}~\bibnamefont {Jiang}},
  \bibinfo {author} {\bibfnamefont {F.}~\bibnamefont {Schedin}}, \bibinfo
  {author} {\bibfnamefont {T.~J.}\ \bibnamefont {Booth}}, \bibinfo {author}
  {\bibfnamefont {V.}~\bibnamefont {Khotkevich}}, \bibinfo {author}
  {\bibfnamefont {S.~V.}\ \bibnamefont {Morozov}}, \ and\ \bibinfo {author}
  {\bibfnamefont {A.~K.}\ \bibnamefont {Geim}},\ }\href@noop {} {\bibfield
  {journal} {\bibinfo  {journal} {Proc. Nat. Acc. Sci.}\ }\textbf {\bibinfo
  {volume} {102}},\ \bibinfo {pages} {10451} (\bibinfo {year}
  {2005})}\BibitemShut {NoStop}%
\bibitem [{\citenamefont {Xu}\ \emph {et~al.}(2014)\citenamefont {Xu},
  \citenamefont {Yao}, \citenamefont {Xiao},\ and\ \citenamefont
  {Heinz}}]{Xu2014}%
  \BibitemOpen
  \bibfield  {author} {\bibinfo {author} {\bibfnamefont {X.}~\bibnamefont
  {Xu}}, \bibinfo {author} {\bibfnamefont {W.}~\bibnamefont {Yao}}, \bibinfo
  {author} {\bibfnamefont {D.}~\bibnamefont {Xiao}}, \ and\ \bibinfo {author}
  {\bibfnamefont {T.-F.}\ \bibnamefont {Heinz}},\ }\href@noop {} {\bibfield
  {journal} {\bibinfo  {journal} {Nat. Phys.}\ }\textbf {\bibinfo {volume}
  {10}},\ \bibinfo {pages} {343} (\bibinfo {year} {2014})}\BibitemShut
  {NoStop}%
\bibitem [{\citenamefont {Zhang}\ \emph {et~al.}(2014)\citenamefont {Zhang},
  \citenamefont {Oka}, \citenamefont {Suzuki}, \citenamefont {Ye},\ and\
  \citenamefont {Iwasa}}]{Zhang2014}%
  \BibitemOpen
  \bibfield  {author} {\bibinfo {author} {\bibfnamefont {Y.~J.}\ \bibnamefont
  {Zhang}}, \bibinfo {author} {\bibfnamefont {T.}~\bibnamefont {Oka}}, \bibinfo
  {author} {\bibfnamefont {R.}~\bibnamefont {Suzuki}}, \bibinfo {author}
  {\bibfnamefont {J.~T.}\ \bibnamefont {Ye}}, \ and\ \bibinfo {author}
  {\bibfnamefont {Y.}~\bibnamefont {Iwasa}},\ }\href@noop {} {\bibfield
  {journal} {\bibinfo  {journal} {Science}\ }\textbf {\bibinfo {volume}
  {344}},\ \bibinfo {pages} {725} (\bibinfo {year} {2014})}\BibitemShut
  {NoStop}%
\bibitem [{\citenamefont {Ye}\ \emph {et~al.}(2012)\citenamefont {Ye},
  \citenamefont {Zhang}, \citenamefont {Akashi}, \citenamefont {Bahramy},
  \citenamefont {Arita},\ and\ \citenamefont {Iwasa}}]{Ye2012}%
  \BibitemOpen
  \bibfield  {author} {\bibinfo {author} {\bibfnamefont {J.~T.}\ \bibnamefont
  {Ye}}, \bibinfo {author} {\bibfnamefont {Y.~J.}\ \bibnamefont {Zhang}},
  \bibinfo {author} {\bibfnamefont {R.}~\bibnamefont {Akashi}}, \bibinfo
  {author} {\bibfnamefont {M.~S.}\ \bibnamefont {Bahramy}}, \bibinfo {author}
  {\bibfnamefont {R.}~\bibnamefont {Arita}}, \ and\ \bibinfo {author}
  {\bibfnamefont {Y.}~\bibnamefont {Iwasa}},\ }\href@noop {} {\bibfield
  {journal} {\bibinfo  {journal} {Science}\ }\textbf {\bibinfo {volume}
  {338}},\ \bibinfo {pages} {1193} (\bibinfo {year} {2012})}\BibitemShut
  {NoStop}%
\bibitem [{\citenamefont {Gregory}\ \emph {et~al.}(1998)\citenamefont
  {Gregory}, \citenamefont {Barker}, \citenamefont {Edwards}, \citenamefont
  {Slaskic},\ and\ \citenamefont {Siddonsa}}]{Gregory1998}%
  \BibitemOpen
  \bibfield  {author} {\bibinfo {author} {\bibfnamefont {D.}~\bibnamefont
  {Gregory}}, \bibinfo {author} {\bibfnamefont {M.}~\bibnamefont {Barker}},
  \bibinfo {author} {\bibfnamefont {P.}~\bibnamefont {Edwards}}, \bibinfo
  {author} {\bibfnamefont {M.}~\bibnamefont {Slaskic}}, \ and\ \bibinfo
  {author} {\bibfnamefont {D.}~\bibnamefont {Siddonsa}},\ }\href@noop {}
  {\bibfield  {journal} {\bibinfo  {journal} {J. Solid State Chem.}\ }\textbf
  {\bibinfo {volume} {137}},\ \bibinfo {pages} {62} (\bibinfo {year}
  {1998})}\BibitemShut {NoStop}%
\bibitem [{\citenamefont {Yamanaka}\ \emph {et~al.}(1996)\citenamefont
  {Yamanaka}, \citenamefont {Kawaji}, \citenamefont {i.~Hotehama},\ and\
  \citenamefont {Ohashi}}]{Yamanaka1996}%
  \BibitemOpen
  \bibfield  {author} {\bibinfo {author} {\bibfnamefont {S.}~\bibnamefont
  {Yamanaka}}, \bibinfo {author} {\bibfnamefont {H.}~\bibnamefont {Kawaji}},
  \bibinfo {author} {\bibfnamefont {K.}~\bibnamefont {i.~Hotehama}}, \ and\
  \bibinfo {author} {\bibfnamefont {M.}~\bibnamefont {Ohashi}},\ }\href@noop {}
  {\bibfield  {journal} {\bibinfo  {journal} {Adv. Mater.}\ }\textbf {\bibinfo
  {volume} {8}},\ \bibinfo {pages} {771} (\bibinfo {year} {1996})}\BibitemShut
  {NoStop}%
\bibitem [{\citenamefont {Yamanaka}\ \emph {et~al.}(1998)\citenamefont
  {Yamanaka}, \citenamefont {Hotehama},\ and\ \citenamefont
  {Kawaji}}]{Yamanaka1998}%
  \BibitemOpen
  \bibfield  {author} {\bibinfo {author} {\bibfnamefont {S.}~\bibnamefont
  {Yamanaka}}, \bibinfo {author} {\bibfnamefont {K.}~\bibnamefont {Hotehama}},
  \ and\ \bibinfo {author} {\bibfnamefont {H.}~\bibnamefont {Kawaji}},\
  }\href@noop {} {\bibfield  {journal} {\bibinfo  {journal} {Nature (London)}\
  }\textbf {\bibinfo {volume} {392}},\ \bibinfo {pages} {580} (\bibinfo {year}
  {1998})}\BibitemShut {NoStop}%
\bibitem [{\citenamefont {Taguchi}\ \emph {et~al.}(2006)\citenamefont
  {Taguchi}, \citenamefont {Kitora},\ and\ \citenamefont
  {Iwasa}}]{Taguchi2006}%
  \BibitemOpen
  \bibfield  {author} {\bibinfo {author} {\bibfnamefont {Y.}~\bibnamefont
  {Taguchi}}, \bibinfo {author} {\bibfnamefont {A.}~\bibnamefont {Kitora}}, \
  and\ \bibinfo {author} {\bibfnamefont {Y.}~\bibnamefont {Iwasa}},\
  }\href@noop {} {\bibfield  {journal} {\bibinfo  {journal} {Phys. Rev. Lett.}\
  }\textbf {\bibinfo {volume} {97}},\ \bibinfo {pages} {107001} (\bibinfo
  {year} {2006})}\BibitemShut {NoStop}%
\bibitem [{\citenamefont {Takano}\ \emph
  {et~al.}(2008{\natexlab{a}})\citenamefont {Takano}, \citenamefont {Kishiume},
  \citenamefont {Taguchi},\ and\ \citenamefont {Iwasa}}]{Takano2008}%
  \BibitemOpen
  \bibfield  {author} {\bibinfo {author} {\bibfnamefont {T.}~\bibnamefont
  {Takano}}, \bibinfo {author} {\bibfnamefont {T.}~\bibnamefont {Kishiume}},
  \bibinfo {author} {\bibfnamefont {Y.}~\bibnamefont {Taguchi}}, \ and\
  \bibinfo {author} {\bibfnamefont {Y.}~\bibnamefont {Iwasa}},\ }\href@noop {}
  {\bibfield  {journal} {\bibinfo  {journal} {Phys. Rev. Lett.}\ }\textbf
  {\bibinfo {volume} {100}},\ \bibinfo {pages} {247005} (\bibinfo {year}
  {2008}{\natexlab{a}})}\BibitemShut {NoStop}%
\bibitem [{\citenamefont {Takano}\ \emph
  {et~al.}(2008{\natexlab{b}})\citenamefont {Takano}, \citenamefont {Kitora},
  \citenamefont {Taguchi},\ and\ \citenamefont {Iwasa}}]{Takano2008b}%
  \BibitemOpen
  \bibfield  {author} {\bibinfo {author} {\bibfnamefont {T.}~\bibnamefont
  {Takano}}, \bibinfo {author} {\bibfnamefont {A.}~\bibnamefont {Kitora}},
  \bibinfo {author} {\bibfnamefont {Y.}~\bibnamefont {Taguchi}}, \ and\
  \bibinfo {author} {\bibfnamefont {Y.}~\bibnamefont {Iwasa}},\ }\href@noop {}
  {\bibfield  {journal} {\bibinfo  {journal} {Journal of Physics and Chemistry
  of Solids}\ }\textbf {\bibinfo {volume} {69}},\ \bibinfo {pages} {3089}
  (\bibinfo {year} {2008}{\natexlab{b}})}\BibitemShut {NoStop}%
\bibitem [{\citenamefont {Yamanaka}\ \emph {et~al.}(2009)\citenamefont
  {Yamanaka}, \citenamefont {Yasunaga}, \citenamefont {Yamaguchi},\ and\
  \citenamefont {Tagawa}}]{Yamanaka2009}%
  \BibitemOpen
  \bibfield  {author} {\bibinfo {author} {\bibfnamefont {S.}~\bibnamefont
  {Yamanaka}}, \bibinfo {author} {\bibfnamefont {T.}~\bibnamefont {Yasunaga}},
  \bibinfo {author} {\bibfnamefont {K.}~\bibnamefont {Yamaguchi}}, \ and\
  \bibinfo {author} {\bibfnamefont {M.}~\bibnamefont {Tagawa}},\ }\href@noop {}
  {\bibfield  {journal} {\bibinfo  {journal} {J. Mater. Chem.}\ }\textbf
  {\bibinfo {volume} {19}},\ \bibinfo {pages} {2573} (\bibinfo {year}
  {2009})}\BibitemShut {NoStop}%
\bibitem [{\citenamefont {Ye}\ \emph {et~al.}(2010)\citenamefont {Ye},
  \citenamefont {Inoue}, \citenamefont {Kobayashi}, \citenamefont {Kasahara},
  \citenamefont {Yuan}, \citenamefont {Shimotani},\ and\ \citenamefont
  {Iwasa}}]{Ye2010}%
  \BibitemOpen
  \bibfield  {author} {\bibinfo {author} {\bibfnamefont {J.~T.}\ \bibnamefont
  {Ye}}, \bibinfo {author} {\bibfnamefont {S.}~\bibnamefont {Inoue}}, \bibinfo
  {author} {\bibfnamefont {K.}~\bibnamefont {Kobayashi}}, \bibinfo {author}
  {\bibfnamefont {Y.}~\bibnamefont {Kasahara}}, \bibinfo {author}
  {\bibfnamefont {H.~T.}\ \bibnamefont {Yuan}}, \bibinfo {author}
  {\bibfnamefont {H.}~\bibnamefont {Shimotani}}, \ and\ \bibinfo {author}
  {\bibfnamefont {Y.}~\bibnamefont {Iwasa}},\ }\href@noop {} {\bibfield
  {journal} {\bibinfo  {journal} {Nat. Mater.}\ }\textbf {\bibinfo {volume}
  {9}},\ \bibinfo {pages} {125} (\bibinfo {year} {2010})}\BibitemShut {NoStop}%
\bibitem [{\citenamefont {Kasahara}\ \emph {et~al.}(2011)\citenamefont
  {Kasahara}, \citenamefont {Nishijima}, \citenamefont {Sato}, \citenamefont
  {Takeuchi}, \citenamefont {Ye}, \citenamefont {Yuan}, \citenamefont
  {Shimotani},\ and\ \citenamefont {Iwasa}}]{Kasahara2011}%
  \BibitemOpen
  \bibfield  {author} {\bibinfo {author} {\bibfnamefont {Y.}~\bibnamefont
  {Kasahara}}, \bibinfo {author} {\bibfnamefont {T.}~\bibnamefont {Nishijima}},
  \bibinfo {author} {\bibfnamefont {T.}~\bibnamefont {Sato}}, \bibinfo {author}
  {\bibfnamefont {Y.}~\bibnamefont {Takeuchi}}, \bibinfo {author}
  {\bibfnamefont {J.}~\bibnamefont {Ye}}, \bibinfo {author} {\bibfnamefont
  {H.}~\bibnamefont {Yuan}}, \bibinfo {author} {\bibfnamefont {H.}~\bibnamefont
  {Shimotani}}, \ and\ \bibinfo {author} {\bibfnamefont {Y.}~\bibnamefont
  {Iwasa}},\ }\href@noop {} {\bibfield  {journal} {\bibinfo  {journal} {J.
  Phys. Soc. Jpn.}\ }\textbf {\bibinfo {volume} {80}},\ \bibinfo {pages}
  {023708} (\bibinfo {year} {2011})}\BibitemShut {NoStop}%
\bibitem [{\citenamefont {Brumme}\ \emph {et~al.}(2014)\citenamefont {Brumme},
  \citenamefont {Calandra},\ and\ \citenamefont {Mauri}}]{Thomas2014}%
  \BibitemOpen
  \bibfield  {author} {\bibinfo {author} {\bibfnamefont {T.}~\bibnamefont
  {Brumme}}, \bibinfo {author} {\bibfnamefont {M.}~\bibnamefont {Calandra}}, \
  and\ \bibinfo {author} {\bibfnamefont {F.}~\bibnamefont {Mauri}},\
  }\href@noop {} {\bibfield  {journal} {\bibinfo  {journal} {Phys. Rev. B}\
  }\textbf {\bibinfo {volume} {89}},\ \bibinfo {pages} {245406} (\bibinfo
  {year} {2014})}\BibitemShut {NoStop}%
\bibitem [{\citenamefont {Ekimov}\ \emph {et~al.}(2004)\citenamefont {Ekimov},
  \citenamefont {Sidorov}, \citenamefont {Bauer}, \citenamefont {Mel'nik},
  \citenamefont {Curro}, \citenamefont {Thompson},\ and\ \citenamefont
  {Stishov}}]{Ekimov2004}%
  \BibitemOpen
  \bibfield  {author} {\bibinfo {author} {\bibfnamefont {E.~A.}\ \bibnamefont
  {Ekimov}}, \bibinfo {author} {\bibfnamefont {V.~A.}\ \bibnamefont {Sidorov}},
  \bibinfo {author} {\bibfnamefont {E.~D.}\ \bibnamefont {Bauer}}, \bibinfo
  {author} {\bibfnamefont {N.~N.}\ \bibnamefont {Mel'nik}}, \bibinfo {author}
  {\bibfnamefont {N.~J.}\ \bibnamefont {Curro}}, \bibinfo {author}
  {\bibfnamefont {J.~D.}\ \bibnamefont {Thompson}}, \ and\ \bibinfo {author}
  {\bibfnamefont {S.~M.}\ \bibnamefont {Stishov}},\ }\href@noop {} {\bibfield
  {journal} {\bibinfo  {journal} {Nature (London)}\ }\textbf {\bibinfo {volume}
  {428}},\ \bibinfo {pages} {542} (\bibinfo {year} {2004})}\BibitemShut
  {NoStop}%
\bibitem [{\citenamefont {Heid}\ and\ \citenamefont {Bohnen}(2005)}]{Heid2005}%
  \BibitemOpen
  \bibfield  {author} {\bibinfo {author} {\bibfnamefont {R.}~\bibnamefont
  {Heid}}\ and\ \bibinfo {author} {\bibfnamefont {K.-P.}\ \bibnamefont
  {Bohnen}},\ }\href@noop {} {\bibfield  {journal} {\bibinfo  {journal} {Phys.
  Rev. B}\ }\textbf {\bibinfo {volume} {72}},\ \bibinfo {pages} {134527}
  (\bibinfo {year} {2005})}\BibitemShut {NoStop}%
\bibitem [{\citenamefont {Takano}\ \emph {et~al.}(2011)\citenamefont {Takano},
  \citenamefont {Kasahara}, \citenamefont {Oguchi}, \citenamefont {Hase},
  \citenamefont {Taguchi},\ and\ \citenamefont {Iwasa}}]{Takano2011}%
  \BibitemOpen
  \bibfield  {author} {\bibinfo {author} {\bibfnamefont {T.}~\bibnamefont
  {Takano}}, \bibinfo {author} {\bibfnamefont {Y.}~\bibnamefont {Kasahara}},
  \bibinfo {author} {\bibfnamefont {T.}~\bibnamefont {Oguchi}}, \bibinfo
  {author} {\bibfnamefont {I.}~\bibnamefont {Hase}}, \bibinfo {author}
  {\bibfnamefont {Y.}~\bibnamefont {Taguchi}}, \ and\ \bibinfo {author}
  {\bibfnamefont {Y.}~\bibnamefont {Iwasa}},\ }\href@noop {} {\bibfield
  {journal} {\bibinfo  {journal} {J. Phys. Soc. Jpn.}\ }\textbf {\bibinfo
  {volume} {80}},\ \bibinfo {pages} {023702} (\bibinfo {year}
  {2011})}\BibitemShut {NoStop}%
\bibitem [{\citenamefont {Botana}\ and\ \citenamefont
  {Pickett}(2014)}]{Botana2014}%
  \BibitemOpen
  \bibfield  {author} {\bibinfo {author} {\bibfnamefont {A.~S.}\ \bibnamefont
  {Botana}}\ and\ \bibinfo {author} {\bibfnamefont {W.~E.}\ \bibnamefont
  {Pickett}},\ }\href@noop {} {\bibfield  {journal} {\bibinfo  {journal} {Phys.
  Rev. B}\ }\textbf {\bibinfo {volume} {90}},\ \bibinfo {pages} {125145}
  (\bibinfo {year} {2014})}\BibitemShut {NoStop}%
\bibitem [{\citenamefont {Calandra}\ \emph {et~al.}(2015)\citenamefont
  {Calandra}, \citenamefont {Zoccante},\ and\ \citenamefont
  {Mauri}}]{Paolo2015}%
  \BibitemOpen
  \bibfield  {author} {\bibinfo {author} {\bibfnamefont {M.}~\bibnamefont
  {Calandra}}, \bibinfo {author} {\bibfnamefont {P.}~\bibnamefont {Zoccante}},
  \ and\ \bibinfo {author} {\bibfnamefont {F.}~\bibnamefont {Mauri}},\
  }\href@noop {} {\bibfield  {journal} {\bibinfo  {journal} {Phys. Rev. Lett.}\
  }\textbf {\bibinfo {volume} {114}},\ \bibinfo {pages} {077001} (\bibinfo
  {year} {2015})}\BibitemShut {NoStop}%
\bibitem [{\citenamefont {Giuliani}\ and\ \citenamefont
  {Vignale}(2005)}]{GiulianiVignale}%
  \BibitemOpen
  \bibfield  {author} {\bibinfo {author} {\bibfnamefont {G.~F.}\ \bibnamefont
  {Giuliani}}\ and\ \bibinfo {author} {\bibfnamefont {G.}~\bibnamefont
  {Vignale}},\ }\href@noop {} {\emph {\bibinfo {title} {Quantum Theory of the
  Electron Liquid}}}\ (\bibinfo  {publisher} {Cambridge},\ \bibinfo {year}
  {2005})\BibitemShut {NoStop}%
\bibitem [{\citenamefont {Kasahara}\ \emph {et~al.}(2009)\citenamefont
  {Kasahara}, \citenamefont {Kishiume}, \citenamefont {Takano}, \citenamefont
  {Kobayashi}, \citenamefont {Matsuoka}, \citenamefont {Onodera}, \citenamefont
  {Kuroki}, \citenamefont {Taguchi},\ and\ \citenamefont
  {Iwasa}}]{Kasahara2009}%
  \BibitemOpen
  \bibfield  {author} {\bibinfo {author} {\bibfnamefont {Y.}~\bibnamefont
  {Kasahara}}, \bibinfo {author} {\bibfnamefont {T.}~\bibnamefont {Kishiume}},
  \bibinfo {author} {\bibfnamefont {T.}~\bibnamefont {Takano}}, \bibinfo
  {author} {\bibfnamefont {K.}~\bibnamefont {Kobayashi}}, \bibinfo {author}
  {\bibfnamefont {E.}~\bibnamefont {Matsuoka}}, \bibinfo {author}
  {\bibfnamefont {H.}~\bibnamefont {Onodera}}, \bibinfo {author} {\bibfnamefont
  {K.}~\bibnamefont {Kuroki}}, \bibinfo {author} {\bibfnamefont
  {Y.}~\bibnamefont {Taguchi}}, \ and\ \bibinfo {author} {\bibfnamefont
  {Y.}~\bibnamefont {Iwasa}},\ }\href@noop {} {\bibfield  {journal} {\bibinfo
  {journal} {Phys. Rev. Lett.}\ }\textbf {\bibinfo {volume} {103}},\ \bibinfo
  {pages} {077004} (\bibinfo {year} {2009})}\BibitemShut {NoStop}%
\bibitem [{\citenamefont {Taguchi}\ \emph {et~al.}(2010)\citenamefont
  {Taguchi}, \citenamefont {Kasahara}, \citenamefont {Kishiume}, \citenamefont
  {Takano}, \citenamefont {Kobayashi}, \citenamefont {Matsuoka}, \citenamefont
  {Onodera}, \citenamefont {Kuroki},\ and\ \citenamefont
  {Iwasa}}]{Taguchi2010}%
  \BibitemOpen
  \bibfield  {author} {\bibinfo {author} {\bibfnamefont {Y.}~\bibnamefont
  {Taguchi}}, \bibinfo {author} {\bibfnamefont {Y.}~\bibnamefont {Kasahara}},
  \bibinfo {author} {\bibfnamefont {T.}~\bibnamefont {Kishiume}}, \bibinfo
  {author} {\bibfnamefont {T.}~\bibnamefont {Takano}}, \bibinfo {author}
  {\bibfnamefont {K.}~\bibnamefont {Kobayashi}}, \bibinfo {author}
  {\bibfnamefont {E.}~\bibnamefont {Matsuoka}}, \bibinfo {author}
  {\bibfnamefont {H.}~\bibnamefont {Onodera}}, \bibinfo {author} {\bibfnamefont
  {K.}~\bibnamefont {Kuroki}}, \ and\ \bibinfo {author} {\bibfnamefont
  {Y.}~\bibnamefont {Iwasa}},\ }\href@noop {} {\bibfield  {journal} {\bibinfo
  {journal} {Physica C: Superconductivity}\ }\textbf {\bibinfo {volume}
  {470}},\ \bibinfo {pages} {S598} (\bibinfo {year} {2010})}\BibitemShut
  {NoStop}%
\bibitem [{\citenamefont {Chen}\ \emph {et~al.}(2001)\citenamefont {Chen},
  \citenamefont {Koiwasaki},\ and\ \citenamefont {Yamanaka}}]{Chen2001}%
  \BibitemOpen
  \bibfield  {author} {\bibinfo {author} {\bibfnamefont {X.}~\bibnamefont
  {Chen}}, \bibinfo {author} {\bibfnamefont {T.}~\bibnamefont {Koiwasaki}}, \
  and\ \bibinfo {author} {\bibfnamefont {S.}~\bibnamefont {Yamanaka}},\
  }\href@noop {} {\bibfield  {journal} {\bibinfo  {journal} {Journal of Solid
  State Chemistry}\ }\textbf {\bibinfo {volume} {159}},\ \bibinfo {pages} {80}
  (\bibinfo {year} {2001})}\BibitemShut {NoStop}%
\bibitem [{\citenamefont {Shamoto}\ \emph {et~al.}(1998)\citenamefont
  {Shamoto}, \citenamefont {Kato}, \citenamefont {Ono}, \citenamefont
  {Miyazaki}, \citenamefont {Ohoyama}, \citenamefont {Ohashi}, \citenamefont
  {Yamaguchi},\ and\ \citenamefont {Kajitani}}]{Shamoto1998}%
  \BibitemOpen
  \bibfield  {author} {\bibinfo {author} {\bibfnamefont {S.}~\bibnamefont
  {Shamoto}}, \bibinfo {author} {\bibfnamefont {T.}~\bibnamefont {Kato}},
  \bibinfo {author} {\bibfnamefont {Y.}~\bibnamefont {Ono}}, \bibinfo {author}
  {\bibfnamefont {Y.}~\bibnamefont {Miyazaki}}, \bibinfo {author}
  {\bibfnamefont {K.}~\bibnamefont {Ohoyama}}, \bibinfo {author} {\bibfnamefont
  {M.}~\bibnamefont {Ohashi}}, \bibinfo {author} {\bibfnamefont
  {Y.}~\bibnamefont {Yamaguchi}}, \ and\ \bibinfo {author} {\bibfnamefont
  {T.}~\bibnamefont {Kajitani}},\ }\href@noop {} {\bibfield  {journal}
  {\bibinfo  {journal} {Physica C}\ }\textbf {\bibinfo {volume} {306}},\
  \bibinfo {pages} {7} (\bibinfo {year} {1998})}\BibitemShut {NoStop}%
\bibitem [{\citenamefont {Kasahara}\ \emph {et~al.}(2010)\citenamefont
  {Kasahara}, \citenamefont {Kishiume}, \citenamefont {Kobayashi},
  \citenamefont {Taguchi},\ and\ \citenamefont {Iwasa}}]{Kasahara2010}%
  \BibitemOpen
  \bibfield  {author} {\bibinfo {author} {\bibfnamefont {Y.}~\bibnamefont
  {Kasahara}}, \bibinfo {author} {\bibfnamefont {T.}~\bibnamefont {Kishiume}},
  \bibinfo {author} {\bibfnamefont {K.}~\bibnamefont {Kobayashi}}, \bibinfo
  {author} {\bibfnamefont {Y.}~\bibnamefont {Taguchi}}, \ and\ \bibinfo
  {author} {\bibfnamefont {Y.}~\bibnamefont {Iwasa}},\ }\href@noop {}
  {\bibfield  {journal} {\bibinfo  {journal} {Phys. Rev. B}\ }\textbf {\bibinfo
  {volume} {82}},\ \bibinfo {pages} {054504} (\bibinfo {year}
  {2010})}\BibitemShut {NoStop}%
\bibitem [{\citenamefont {Dirac}(1930)}]{Dirac1930}%
  \BibitemOpen
  \bibfield  {author} {\bibinfo {author} {\bibfnamefont {P.}~\bibnamefont
  {Dirac}},\ }\href@noop {} {\bibfield  {journal} {\bibinfo  {journal} {Proc.
  Cambridge Phil. Soc.}\ }\textbf {\bibinfo {volume} {26}},\ \bibinfo {pages}
  {376} (\bibinfo {year} {1930})}\BibitemShut {NoStop}%
\bibitem [{\citenamefont {Perdew}\ and\ \citenamefont {Zunger}(1981)}]{PZ}%
  \BibitemOpen
  \bibfield  {author} {\bibinfo {author} {\bibfnamefont {J.~P.}\ \bibnamefont
  {Perdew}}\ and\ \bibinfo {author} {\bibfnamefont {A.}~\bibnamefont
  {Zunger}},\ }\href@noop {} {\bibfield  {journal} {\bibinfo  {journal} {Phys.
  Rev. B}\ }\textbf {\bibinfo {volume} {23}},\ \bibinfo {pages} {5048}
  (\bibinfo {year} {1981})}\BibitemShut {NoStop}%
\bibitem [{\citenamefont {Perdew}\ \emph {et~al.}(1996)\citenamefont {Perdew},
  \citenamefont {Burke},\ and\ \citenamefont {Ernzerhof}}]{PBE}%
  \BibitemOpen
  \bibfield  {author} {\bibinfo {author} {\bibfnamefont {J.~P.}\ \bibnamefont
  {Perdew}}, \bibinfo {author} {\bibfnamefont {K.}~\bibnamefont {Burke}}, \
  and\ \bibinfo {author} {\bibfnamefont {M.}~\bibnamefont {Ernzerhof}},\
  }\href@noop {} {\bibfield  {journal} {\bibinfo  {journal} {Phys. Rev. Lett.}\
  }\textbf {\bibinfo {volume} {77}},\ \bibinfo {pages} {3865} (\bibinfo {year}
  {1996})}\BibitemShut {NoStop}%
\bibitem [{\citenamefont {Becke}(1993)}]{hybrid1}%
  \BibitemOpen
  \bibfield  {author} {\bibinfo {author} {\bibfnamefont {A.~D.}\ \bibnamefont
  {Becke}},\ }\href@noop {} {\bibfield  {journal} {\bibinfo  {journal} {J.
  Chem. Phys.}\ }\textbf {\bibinfo {volume} {98}},\ \bibinfo {pages} {5648}
  (\bibinfo {year} {1993})}\BibitemShut {NoStop}%
\bibitem [{\citenamefont {Lee}\ \emph {et~al.}(1988)\citenamefont {Lee},
  \citenamefont {Yang},\ and\ \citenamefont {Parr}}]{blyp2}%
  \BibitemOpen
  \bibfield  {author} {\bibinfo {author} {\bibfnamefont {C.}~\bibnamefont
  {Lee}}, \bibinfo {author} {\bibfnamefont {W.}~\bibnamefont {Yang}}, \ and\
  \bibinfo {author} {\bibfnamefont {R.~G.}\ \bibnamefont {Parr}},\ }\href@noop
  {} {\bibfield  {journal} {\bibinfo  {journal} {Phys. Rev. B}\ }\textbf
  {\bibinfo {volume} {37}},\ \bibinfo {pages} {785} (\bibinfo {year}
  {1988})}\BibitemShut {NoStop}%
\bibitem [{\citenamefont {Vosko}\ \emph {et~al.}(1980)\citenamefont {Vosko},
  \citenamefont {Wilk},\ and\ \citenamefont {Nusair}}]{b3lyp}%
  \BibitemOpen
  \bibfield  {author} {\bibinfo {author} {\bibfnamefont {S.~H.}\ \bibnamefont
  {Vosko}}, \bibinfo {author} {\bibfnamefont {L.}~\bibnamefont {Wilk}}, \ and\
  \bibinfo {author} {\bibfnamefont {M.}~\bibnamefont {Nusair}},\ }\href@noop {}
  {\bibfield  {journal} {\bibinfo  {journal} {Can. J. Phys.}\ }\textbf
  {\bibinfo {volume} {58}},\ \bibinfo {pages} {1200} (\bibinfo {year}
  {1980})}\BibitemShut {NoStop}%
\bibitem [{\citenamefont {Adamo}\ and\ \citenamefont {Barone}(1999)}]{PBE0}%
  \BibitemOpen
  \bibfield  {author} {\bibinfo {author} {\bibfnamefont {C.}~\bibnamefont
  {Adamo}}\ and\ \bibinfo {author} {\bibfnamefont {V.}~\bibnamefont {Barone}},\
  }\href@noop {} {\bibfield  {journal} {\bibinfo  {journal} {J. Chem. Phys.}\
  }\textbf {\bibinfo {volume} {110}},\ \bibinfo {pages} {6158} (\bibinfo {year}
  {1999})}\BibitemShut {NoStop}%
\bibitem [{\citenamefont {Krukau}\ \emph {et~al.}(2006)\citenamefont {Krukau},
  \citenamefont {Vydrov}, \citenamefont {Izmaylov},\ and\ \citenamefont
  {Scuseria}}]{HSE06}%
  \BibitemOpen
  \bibfield  {author} {\bibinfo {author} {\bibfnamefont {A.~V.}\ \bibnamefont
  {Krukau}}, \bibinfo {author} {\bibfnamefont {O.~A.}\ \bibnamefont {Vydrov}},
  \bibinfo {author} {\bibfnamefont {A.~F.}\ \bibnamefont {Izmaylov}}, \ and\
  \bibinfo {author} {\bibfnamefont {G.~E.}\ \bibnamefont {Scuseria}},\
  }\href@noop {} {\bibfield  {journal} {\bibinfo  {journal} {J. Chem. Phys.}\
  }\textbf {\bibinfo {volume} {125}},\ \bibinfo {pages} {224106} (\bibinfo
  {year} {2006})}\BibitemShut {NoStop}%
\bibitem [{\citenamefont {Dovesi}\ \emph
  {et~al.}(2014{\natexlab{a}})\citenamefont {Dovesi}, \citenamefont {Orlando},
  \citenamefont {Erba}, \citenamefont {Zicovich-Wilson}, \citenamefont
  {Civalleri}, \citenamefont {Casassa}, \citenamefont {Maschio}, \citenamefont
  {Ferrabone}, \citenamefont {Pierre}, \citenamefont {D¿Arco}, \citenamefont
  {Noel}, \citenamefont {Causa}, \citenamefont {Rerat},\ and\ \citenamefont
  {Kirtman}}]{Crystal14}%
  \BibitemOpen
  \bibfield  {author} {\bibinfo {author} {\bibfnamefont {R.}~\bibnamefont
  {Dovesi}}, \bibinfo {author} {\bibfnamefont {R.}~\bibnamefont {Orlando}},
  \bibinfo {author} {\bibfnamefont {A.}~\bibnamefont {Erba}}, \bibinfo {author}
  {\bibfnamefont {C.~M.}\ \bibnamefont {Zicovich-Wilson}}, \bibinfo {author}
  {\bibfnamefont {B.}~\bibnamefont {Civalleri}}, \bibinfo {author}
  {\bibfnamefont {S.}~\bibnamefont {Casassa}}, \bibinfo {author} {\bibfnamefont
  {L.}~\bibnamefont {Maschio}}, \bibinfo {author} {\bibfnamefont
  {M.}~\bibnamefont {Ferrabone}}, \bibinfo {author} {\bibfnamefont {M.~D.~L.}\
  \bibnamefont {Pierre}}, \bibinfo {author} {\bibfnamefont {P.}~\bibnamefont
  {D¿Arco}}, \bibinfo {author} {\bibfnamefont {Y.}~\bibnamefont {Noel}},
  \bibinfo {author} {\bibfnamefont {M.}~\bibnamefont {Causa}}, \bibinfo
  {author} {\bibfnamefont {M.}~\bibnamefont {Rerat}}, \ and\ \bibinfo {author}
  {\bibfnamefont {B.}~\bibnamefont {Kirtman}},\ }\href@noop {} {\bibfield
  {journal} {\bibinfo  {journal} {Int. J. Quantum Chem.}\ }\textbf {\bibinfo
  {volume} {114}},\ \bibinfo {pages} {1287} (\bibinfo {year}
  {2014}{\natexlab{a}})}\BibitemShut {NoStop}%
\bibitem [{\citenamefont {Weigend}\ and\ \citenamefont
  {Ahlrichs}(2005)}]{TZVPbasis}%
  \BibitemOpen
  \bibfield  {author} {\bibinfo {author} {\bibfnamefont {F.}~\bibnamefont
  {Weigend}}\ and\ \bibinfo {author} {\bibfnamefont {R.}~\bibnamefont
  {Ahlrichs}},\ }\href@noop {} {\bibfield  {journal} {\bibinfo  {journal}
  {Phys. Chem. Chem. Phys.}\ }\textbf {\bibinfo {volume} {7}},\ \bibinfo
  {pages} {3297} (\bibinfo {year} {2005})}\BibitemShut {NoStop}%
\bibitem [{\citenamefont {Giannozzi}\ \emph {et~al.}(2009)\citenamefont
  {Giannozzi}, \citenamefont {Baroni}, \citenamefont {Bonini}, \citenamefont
  {Calandra}, \citenamefont {Car}, \citenamefont {Cavazzoni}, \citenamefont
  {Ceresoli}, \citenamefont {Chiarotti}, \citenamefont {Cococcioni},
  \citenamefont {Dabo}, \citenamefont {{Dal Corso}}, \citenamefont
  {de~Gironcoli}, \citenamefont {Fabris}, \citenamefont {Fratesi},
  \citenamefont {Gebauer}, \citenamefont {Gerstmann}, \citenamefont
  {Gougoussis}, \citenamefont {Kokalj}, \citenamefont {Lazzeri}, \citenamefont
  {Martin-Samos}, \citenamefont {Marzari}, \citenamefont {Mauri}, \citenamefont
  {Mazzarello}, \citenamefont {Paolini}, \citenamefont {Pasquarello},
  \citenamefont {Paulatto}, \citenamefont {Sbraccia}, \citenamefont {Scandolo},
  \citenamefont {Sclauzero}, \citenamefont {Seitsonen}, \citenamefont
  {Smogunov}, \citenamefont {Umari},\ and\ \citenamefont {Wentzcovitch}}]{QE}%
  \BibitemOpen
  \bibfield  {author} {\bibinfo {author} {\bibfnamefont {P.}~\bibnamefont
  {Giannozzi}}, \bibinfo {author} {\bibfnamefont {S.}~\bibnamefont {Baroni}},
  \bibinfo {author} {\bibfnamefont {N.}~\bibnamefont {Bonini}}, \bibinfo
  {author} {\bibfnamefont {M.}~\bibnamefont {Calandra}}, \bibinfo {author}
  {\bibfnamefont {R.}~\bibnamefont {Car}}, \bibinfo {author} {\bibfnamefont
  {C.}~\bibnamefont {Cavazzoni}}, \bibinfo {author} {\bibfnamefont
  {D.}~\bibnamefont {Ceresoli}}, \bibinfo {author} {\bibfnamefont {G.~L.}\
  \bibnamefont {Chiarotti}}, \bibinfo {author} {\bibfnamefont {M.}~\bibnamefont
  {Cococcioni}}, \bibinfo {author} {\bibfnamefont {I.}~\bibnamefont {Dabo}},
  \bibinfo {author} {\bibfnamefont {A.}~\bibnamefont {{Dal Corso}}}, \bibinfo
  {author} {\bibfnamefont {S.}~\bibnamefont {de~Gironcoli}}, \bibinfo {author}
  {\bibfnamefont {S.}~\bibnamefont {Fabris}}, \bibinfo {author} {\bibfnamefont
  {G.}~\bibnamefont {Fratesi}}, \bibinfo {author} {\bibfnamefont
  {R.}~\bibnamefont {Gebauer}}, \bibinfo {author} {\bibfnamefont
  {U.}~\bibnamefont {Gerstmann}}, \bibinfo {author} {\bibfnamefont
  {C.}~\bibnamefont {Gougoussis}}, \bibinfo {author} {\bibfnamefont
  {A.}~\bibnamefont {Kokalj}}, \bibinfo {author} {\bibfnamefont
  {M.}~\bibnamefont {Lazzeri}}, \bibinfo {author} {\bibfnamefont
  {L.}~\bibnamefont {Martin-Samos}}, \bibinfo {author} {\bibfnamefont
  {N.}~\bibnamefont {Marzari}}, \bibinfo {author} {\bibfnamefont
  {F.}~\bibnamefont {Mauri}}, \bibinfo {author} {\bibfnamefont
  {R.}~\bibnamefont {Mazzarello}}, \bibinfo {author} {\bibfnamefont
  {S.}~\bibnamefont {Paolini}}, \bibinfo {author} {\bibfnamefont
  {A.}~\bibnamefont {Pasquarello}}, \bibinfo {author} {\bibfnamefont
  {L.}~\bibnamefont {Paulatto}}, \bibinfo {author} {\bibfnamefont
  {C.}~\bibnamefont {Sbraccia}}, \bibinfo {author} {\bibfnamefont
  {S.}~\bibnamefont {Scandolo}}, \bibinfo {author} {\bibfnamefont
  {G.}~\bibnamefont {Sclauzero}}, \bibinfo {author} {\bibfnamefont {A.~P.}\
  \bibnamefont {Seitsonen}}, \bibinfo {author} {\bibfnamefont {A.}~\bibnamefont
  {Smogunov}}, \bibinfo {author} {\bibfnamefont {P.}~\bibnamefont {Umari}}, \
  and\ \bibinfo {author} {\bibfnamefont {R.~M.}\ \bibnamefont {Wentzcovitch}},\
  }\href {http://www.quantum-espresso.org} {\bibfield  {journal} {\bibinfo
  {journal} {J. Phys. Condens. Matter}\ }\textbf {\bibinfo {volume} {21}},\
  \bibinfo {pages} {395502} (\bibinfo {year} {2009})}\BibitemShut {NoStop}%
\bibitem [{\citenamefont {Dovesi}\ \emph
  {et~al.}(2014{\natexlab{b}})\citenamefont {Dovesi}, \citenamefont {Saunders},
  \citenamefont {Roetti}, \citenamefont {Orlando}, \citenamefont
  {Zicovich-Wilson}, \citenamefont {Pascale}, \citenamefont {Doll},
  \citenamefont {Harrison}, \citenamefont {Civalleri}, \citenamefont {Bush},
  \citenamefont {D'Arco}, \citenamefont {Llunell}, \citenamefont {Caus\`{a}},\
  and\ \citenamefont {No\"{e}l}}]{manualcrystal14}%
  \BibitemOpen
  \bibfield  {author} {\bibinfo {author} {\bibfnamefont {R.}~\bibnamefont
  {Dovesi}}, \bibinfo {author} {\bibfnamefont {V.~R.}\ \bibnamefont
  {Saunders}}, \bibinfo {author} {\bibfnamefont {C.}~\bibnamefont {Roetti}},
  \bibinfo {author} {\bibfnamefont {R.}~\bibnamefont {Orlando}}, \bibinfo
  {author} {\bibfnamefont {C.~M.}\ \bibnamefont {Zicovich-Wilson}}, \bibinfo
  {author} {\bibfnamefont {F.}~\bibnamefont {Pascale}}, \bibinfo {author}
  {\bibfnamefont {K.}~\bibnamefont {Doll}}, \bibinfo {author} {\bibfnamefont
  {N.~M.}\ \bibnamefont {Harrison}}, \bibinfo {author} {\bibfnamefont
  {B.}~\bibnamefont {Civalleri}}, \bibinfo {author} {\bibfnamefont {I.~J.}\
  \bibnamefont {Bush}}, \bibinfo {author} {\bibfnamefont {{\relax
  Ph}.}~\bibnamefont {D'Arco}}, \bibinfo {author} {\bibfnamefont
  {M.}~\bibnamefont {Llunell}}, \bibinfo {author} {\bibfnamefont
  {M.}~\bibnamefont {Caus\`{a}}}, \ and\ \bibinfo {author} {\bibfnamefont
  {Y.}~\bibnamefont {No\"{e}l}},\ }\href@noop {} {\emph {\bibinfo {title}
  {\texttt{CRYSTAL14} User's Manual}}},\ \bibinfo {address} {Universit\`{a} di
  Torino, Torino} (\bibinfo {year} {2014}{\natexlab{b}}),\ \bibinfo {note}
  {http://www.crystal.unito.it}\BibitemShut {NoStop}%
\bibitem [{\citenamefont {Ohashi}\ \emph {et~al.}(1989)\citenamefont {Ohashi},
  \citenamefont {Yamanaka},\ and\ \citenamefont {Hattori}}]{Ohashi1989}%
  \BibitemOpen
  \bibfield  {author} {\bibinfo {author} {\bibfnamefont {M.}~\bibnamefont
  {Ohashi}}, \bibinfo {author} {\bibfnamefont {S.}~\bibnamefont {Yamanaka}}, \
  and\ \bibinfo {author} {\bibfnamefont {M.}~\bibnamefont {Hattori}},\
  }\href@noop {} {\bibfield  {journal} {\bibinfo  {journal} {Journal of the
  Ceramic Society of Japan}\ }\textbf {\bibinfo {volume} {97}},\ \bibinfo
  {pages} {1181} (\bibinfo {year} {1989})}\BibitemShut {NoStop}%
\bibitem [{\citenamefont {Yokoya}\ \emph {et~al.}(2004)\citenamefont {Yokoya},
  \citenamefont {Takeuchi}, \citenamefont {Tsuda}, \citenamefont {Kiss},
  \citenamefont {Higuchi}, \citenamefont {Shin}, \citenamefont {Iizawa},
  \citenamefont {Shamoto}, \citenamefont {Kajitani},\ and\ \citenamefont
  {Takahashi}}]{Yokoya2004}%
  \BibitemOpen
  \bibfield  {author} {\bibinfo {author} {\bibfnamefont {T.}~\bibnamefont
  {Yokoya}}, \bibinfo {author} {\bibfnamefont {T.}~\bibnamefont {Takeuchi}},
  \bibinfo {author} {\bibfnamefont {S.}~\bibnamefont {Tsuda}}, \bibinfo
  {author} {\bibfnamefont {T.}~\bibnamefont {Kiss}}, \bibinfo {author}
  {\bibfnamefont {T.}~\bibnamefont {Higuchi}}, \bibinfo {author} {\bibfnamefont
  {S.}~\bibnamefont {Shin}}, \bibinfo {author} {\bibfnamefont {K.}~\bibnamefont
  {Iizawa}}, \bibinfo {author} {\bibfnamefont {S.}~\bibnamefont {Shamoto}},
  \bibinfo {author} {\bibfnamefont {T.}~\bibnamefont {Kajitani}}, \ and\
  \bibinfo {author} {\bibfnamefont {T.}~\bibnamefont {Takahashi}},\ }\href@noop
  {} {\bibfield  {journal} {\bibinfo  {journal} {Phys. Rev. B}\ }\textbf
  {\bibinfo {volume} {70}},\ \bibinfo {pages} {193103} (\bibinfo {year}
  {2004})}\BibitemShut {NoStop}%
\bibitem [{\citenamefont {Marchi}\ \emph {et~al.}(2009)\citenamefont {Marchi},
  \citenamefont {De~Palo}, \citenamefont {Moroni},\ and\ \citenamefont
  {Senatore}}]{Marchi2009}%
  \BibitemOpen
  \bibfield  {author} {\bibinfo {author} {\bibfnamefont {M.}~\bibnamefont
  {Marchi}}, \bibinfo {author} {\bibfnamefont {S.}~\bibnamefont {De~Palo}},
  \bibinfo {author} {\bibfnamefont {S.}~\bibnamefont {Moroni}}, \ and\ \bibinfo
  {author} {\bibfnamefont {G.}~\bibnamefont {Senatore}},\ }\href@noop {}
  {\bibfield  {journal} {\bibinfo  {journal} {Phys. Rev. B}\ }\textbf {\bibinfo
  {volume} {80}},\ \bibinfo {pages} {035103} (\bibinfo {year}
  {2009})}\BibitemShut {NoStop}%
\bibitem [{\citenamefont {Zhang}\ and\ \citenamefont
  {Das~Sarma}(2005)}]{DasSarma2005}%
  \BibitemOpen
  \bibfield  {author} {\bibinfo {author} {\bibfnamefont {Y.}~\bibnamefont
  {Zhang}}\ and\ \bibinfo {author} {\bibfnamefont {S.}~\bibnamefont
  {Das~Sarma}},\ }\href@noop {} {\bibfield  {journal} {\bibinfo  {journal}
  {Phys. Rev. B}\ }\textbf {\bibinfo {volume} {72}},\ \bibinfo {pages} {075308}
  (\bibinfo {year} {2005})}\BibitemShut {NoStop}%
\bibitem [{\citenamefont {Kaur}\ \emph {et~al.}(2010)\citenamefont {Kaur},
  \citenamefont {Ylvisaker}, \citenamefont {Li}, \citenamefont {Galli},\ and\
  \citenamefont {Pickett}}]{GalliPickett2010}%
  \BibitemOpen
  \bibfield  {author} {\bibinfo {author} {\bibfnamefont {A.}~\bibnamefont
  {Kaur}}, \bibinfo {author} {\bibfnamefont {E.~R.}\ \bibnamefont {Ylvisaker}},
  \bibinfo {author} {\bibfnamefont {Y.}~\bibnamefont {Li}}, \bibinfo {author}
  {\bibfnamefont {G.}~\bibnamefont {Galli}}, \ and\ \bibinfo {author}
  {\bibfnamefont {W.~E.}\ \bibnamefont {Pickett}},\ }\href@noop {} {\bibfield
  {journal} {\bibinfo  {journal} {Phys. Rev. B}\ }\textbf {\bibinfo {volume}
  {82}},\ \bibinfo {pages} {155125} (\bibinfo {year} {2010})}\BibitemShut
  {NoStop}%
\bibitem [{\citenamefont {Adelmann}\ \emph {et~al.}(1999)\citenamefont
  {Adelmann}, \citenamefont {Renker}, \citenamefont {Schober}, \citenamefont
  {Braden},\ and\ \citenamefont {Fernandez-Dias}}]{Adelmann1999}%
  \BibitemOpen
  \bibfield  {author} {\bibinfo {author} {\bibfnamefont {P.}~\bibnamefont
  {Adelmann}}, \bibinfo {author} {\bibfnamefont {B.}~\bibnamefont {Renker}},
  \bibinfo {author} {\bibfnamefont {H.}~\bibnamefont {Schober}}, \bibinfo
  {author} {\bibfnamefont {M.}~\bibnamefont {Braden}}, \ and\ \bibinfo {author}
  {\bibfnamefont {F.}~\bibnamefont {Fernandez-Dias}},\ }\href@noop {}
  {\bibfield  {journal} {\bibinfo  {journal} {J. Low Temp. Phys.}\ }\textbf
  {\bibinfo {volume} {117}},\ \bibinfo {pages} {449} (\bibinfo {year}
  {1999})}\BibitemShut {NoStop}%
\bibitem [{\citenamefont {Cros}\ \emph {et~al.}(2003)\citenamefont {Cros},
  \citenamefont {Cantarero}, \citenamefont {Beltr\'{a}n-Porter}, \citenamefont
  {Or\'{o}-Sol\'{e}},\ and\ \citenamefont {Fuertes}}]{Cros2003}%
  \BibitemOpen
  \bibfield  {author} {\bibinfo {author} {\bibfnamefont {A.}~\bibnamefont
  {Cros}}, \bibinfo {author} {\bibfnamefont {A.}~\bibnamefont {Cantarero}},
  \bibinfo {author} {\bibfnamefont {D.}~\bibnamefont {Beltr\'{a}n-Porter}},
  \bibinfo {author} {\bibfnamefont {J.}~\bibnamefont {Or\'{o}-Sol\'{e}}}, \
  and\ \bibinfo {author} {\bibfnamefont {A.}~\bibnamefont {Fuertes}},\
  }\href@noop {} {\bibfield  {journal} {\bibinfo  {journal} {Phys. Rev. B}\
  }\textbf {\bibinfo {volume} {67}},\ \bibinfo {pages} {104502} (\bibinfo
  {year} {2003})}\BibitemShut {NoStop}%
\bibitem [{\citenamefont {Kitora}\ \emph {et~al.}(2007)\citenamefont {Kitora},
  \citenamefont {Taguchi},\ and\ \citenamefont {Iwasa}}]{Kitora2007}%
  \BibitemOpen
  \bibfield  {author} {\bibinfo {author} {\bibfnamefont {A.}~\bibnamefont
  {Kitora}}, \bibinfo {author} {\bibfnamefont {Y.}~\bibnamefont {Taguchi}}, \
  and\ \bibinfo {author} {\bibfnamefont {Y.}~\bibnamefont {Iwasa}},\
  }\href@noop {} {\bibfield  {journal} {\bibinfo  {journal} {J. Phys. Soc.
  Jpn.}\ }\textbf {\bibinfo {volume} {76}},\ \bibinfo {pages} {023706}
  (\bibinfo {year} {2007})}\BibitemShut {NoStop}%
\end{thebibliography}%

\end{document}